\documentclass[12pt]{article}
\usepackage{preamble}
\usepackage{natbib}
\usepackage{graphicx}
\usepackage[english]{babel}
\usepackage[utf8]{inputenc}
\usepackage{mathtools}
\usepackage{amsmath,amsthm,amssymb,bbm}
\newtheorem*{theorem*}{Theorem}
\newtheorem*{prop*}{Proposition}
\newcommand*\samethanks[1][\value{footnote}]{\footnotemark[#1]}

\title{Bayesian Optimization Sequential Surrogate (BOSS) Algorithm: Fast Bayesian Inference for a Broad Class of Bayesian Hierarchical Models}
\date{} 					

\author{
  Dayi Li\thanks{These authors contributed equally to this work. The names are listed alphabetically.} \\
  Department of Statistical Sciences, University of Toronto \\
  \and
  Ziang Zhang\samethanks\  \thanks{Corresponding author: aguero.zhang@mail.utoronto.ca} \\
  Department of Statistical Sciences, University of Toronto\\
}

\begin{document}
\maketitle
\begin{abstract}
    Approximate Bayesian inference based on Laplace approximation and quadrature methods have become increasingly popular for their efficiency at fitting latent Gaussian models (LGM), which encompass popular models such as Bayesian generalized linear models, survival models, and spatio-temporal models.
    However, many useful models fall under the LGM framework only if some conditioning parameters are fixed, as the design matrix would vary with these parameters otherwise.
    Such models are termed the conditional LGMs with examples in change-point detection, non-linear regression, etc.
    Existing methods for fitting conditional LGMs rely on grid search or Markov-chain Monte Carlo (MCMC); both require a large number of evaluations of the unnormalized posterior density of the conditioning parameters. 
    As each evaluation of the density requires fitting a separate LGM, these methods become computationally prohibitive beyond simple scenarios.
    In this work, we introduce the Bayesian optimization sequential surrogate (BOSS) algorithm, which combines Bayesian optimization with approximate Bayesian inference methods to significantly reduce the computational resources required for fitting conditional LGMs.
    With orders of magnitude fewer evaluations compared to grid or MCMC methods, Bayesian optimization provides us with sequential design points that capture the majority of the posterior mass of the conditioning parameters, which subsequently yields an accurate surrogate posterior distribution that can be easily normalized.
    We illustrate the efficiency, accuracy, and practical utility of the proposed method through extensive simulation studies and real-world applications in epidemiology, environmental sciences, and astrophysics.
\end{abstract}
\noindent%

\doublespacing

\section{Introduction}\label{sec:intro}

Making efficient inference of complex hierarchical models has been a central theme in Bayesian statistics for a long time.
For a wide class of hierarchical models that belong to the Latent Gaussian Models (LGM) \citep{inla} or the extended Latent Gaussian Models (ELGM) \citep{elgm}, efficient approximate inference methods have been proposed based on the nested Laplace approximations and quadrature methods.
These approximate inference methods have become increasingly popular in diverse applications in epidemiology \citep{knutson2023estimating}, environmental sciences \citep{halonen2016long}, genetics \citep{niemi2015empirical} and astrophysics \citep{Li_2022}; 
see \cite{rue2017bayesian} and \cite{van2023new} for exhaustive lists of such examples.
Their theoretical guarantees are also well-established in the existing literature \citep{aghqtheory}.

However, many other useful models fall under the LGM (or ELGM) framework only if some parameters, which we call the conditioning parameters, are fixed \citep{bivand2014approximate}, as the design matrix would vary with these parameters otherwise.
Such models are termed conditional LGM, and include models having unknown change points \citep{altieri2015changepoint}, non-linear regression parameters \citep{bachl2019inlabru,bayliss2020data, Li_2022}, or Gaussian process priors approximated by varying finite element approximations \citep{sgp,IwpOsplines}.

Existing methods to fit conditional LGMs rely on grid search or Markov-chain Monte Carlo (MCMC) to explore the posterior density of the conditioning parameters with a large number of evaluations.
\cite{bivand2014approximate} evaluated the density on a regular grid over the support of the conditioning parameter, and combined the fitted LGMs through the Bayesian model averaging using these evaluations.
\cite{gomez2020bayesian} extended the grid approach in \cite{bivand2014approximate} for conditional LGMs with more than one conditioning parameter.
\cite{gomez2018markov} on the other hand, uses MCMC to obtain posterior samples of the conditioning parameters, where each MCMC iteration evaluates the density.
These approaches provide solutions to fit some simple conditional LGMs, but become computationally infeasible as the model structure gets complicated.

In this paper, we propose an efficient approximate inference approach for conditional LGMs, which requires only a small number of evaluations of the (unnormalized) posterior density of the conditioning parameter.
Our approach sequentially select a series of design points of the conditioning parameters that capture the majority of its posterior mass through a Bayesian Optimization (BO) procedure, and then construct a representative surrogate function for the posterior using the design points. 
Consequently, we term the method the Bayesian Optimization Sequential Surrogate (BOSS) algorithm. 
Through simulations, we show that the BOSS method produces inferential results that are indistinguishable from the existing grid/MCMC methods, with only a fraction of the computational cost.
Finally, we illustrate the practical utility of the proposed BOSS method through various real world applications in epidemiology, environmental science and astrophysics.

The rest of this paper is structured as follows: \cref{sec:prelim} provides a preliminary background on LGMs and ELGMs as well as efficient approximate Bayesian inference methods for fitting them. 
We formally define in \cref{sec:method} the conditional LGM framework and introduce the proposed BOSS algorithm and contains details of applying BOSS for fitting conditional LGMs. \cref{sec:simulation} showcases extensive simulation examples to demonstrate the efficiency and accuracy of our algorithm. \cref{sec:applications} contains multiple real-world problems under the framework of conditional LGMs and we apply the proposed algorithm to make inferences for these problems.
We conclude with summaries and discussions in \cref{sec:discussion}.

\section{Preliminary}\label{sec:prelim}

\subsection{LGMs and ELGMs}\label{subsec:ELGM}

An Extended Latent Gaussian Model (ELGM) takes the following form \citep{elgm}:
\begin{equation}\label{equ:ELGMdef}
    \begin{aligned}
    y_i|\boldsymbol{\eta},\boldsymbol{\theta}_1 &\overset{ind}{\sim} \pi(y_i|\boldsymbol{\eta}, \boldsymbol{\theta}_1), \ \forall i \in [n]  \\
    \eta_i &= \boldsymbol{v}_i ^T \boldsymbol{\beta} + \sum_{l=1}^{L} g_{l}(x_{li}),
    \end{aligned}
\end{equation}
where $n$ denotes the sample size, $\boldsymbol{y} = \{y_i\}_{i=1}^n$ denotes the response variable, $\boldsymbol{v}_i$ and $x_{li}$ denote the covariates. 
$\pi(y_i|\boldsymbol{\eta}, \boldsymbol{\theta}_1)$ is the density function of $y_i$ with linear predictors $\boldsymbol{\eta} = \{\eta_i\}_{i=1}^n$ and hyperparameter $\boldsymbol{\theta}_1$.
The covariate $\boldsymbol{v}$ is assumed to have linear fixed effect $\boldsymbol{\beta}$ assigned with a Gaussian prior $\boldsymbol{\beta} \sim \mathcal{N}(\boldsymbol{0}, \boldsymbol{\Sigma}_\beta)$ and each $g_l$ denotes a Gaussian random effect component. 
To simplify the notation, let $\boldsymbol{g}_l = \left\{g_l(x_{li})\right\}_{i=1}^n$ and $\boldsymbol{G} = \{\boldsymbol{g}_l\}_{l=1}^L$ be the collection of all random components in the model, and assume $\boldsymbol{G} \sim \mathcal{N}\left\{\boldsymbol{0}, \Sigma_{\boldsymbol{G}}(\boldsymbol{\theta}_2)\right\}$, where $\boldsymbol{\theta}_2$ denotes the hyperparameters that characterize the covariance of $\boldsymbol{G}$.

Given the latent field $\boldsymbol{U} = (\boldsymbol{\beta}, \boldsymbol{G})$, the linear predictor can be written as $\boldsymbol{\eta} = \boldsymbol{A} \boldsymbol{U}$ for a fixed design matrix $\boldsymbol{A}$.
The latent field has a Gaussian prior $\boldsymbol{U} \sim \mathcal{N}\left\{\boldsymbol{0}, \boldsymbol{\Sigma}(\boldsymbol{\theta}_2)\right\}$.
When the distribution of each response variable $y_i$ only depends on a single linear predictor $\eta_i$ i.e. $\pi(y_i|\boldsymbol{\eta}, \boldsymbol{\theta}_1) = \pi(y_i|\eta_i, \boldsymbol{\theta}_1)$, the ELGM reduces to a traditional Latent Gaussian Model (LGM) as defined in \cite{inla}.
The prior on the hyperparameter $\boldsymbol{\theta} = (\boldsymbol{\theta}_1, \boldsymbol{\theta}_2)$, however, does not need to be Gaussian. 
For simplicity, we refer to all models under the framework in \cref{equ:ELGMdef} as LGMs, but note that the methods described in this paper apply more broadly to ELGMs as well.

\subsection{Approximate Bayesian Inference}\label{subsec:AGHQ}

For an LGM in \cref{equ:ELGMdef}, efficient posterior approximations for $\pi(\boldsymbol{U}|\boldsymbol{y})$ and $\pi(\boldsymbol{\theta}|\boldsymbol{y})$ have been proposed in \cite{inla,elgm}, through the nested use of Gaussian approximation, Laplace approximation and quadrature methods. 
Given a particular value of $\boldsymbol{\theta}$, the following Gaussian and Laplace approximations are computed:
\[
\widetilde{\pi}_G(\boldsymbol{U}|\boldsymbol{\theta}, \boldsymbol{y}) = d\mathcal{N}\left\{\widehat{\boldsymbol{U}}_{\boldsymbol{\theta}}, \boldsymbol{H}_{\widehat{\boldsymbol{U}}_{\boldsymbol{\theta}}}(\boldsymbol{\theta})^{-1}\right\},\quad
\widetilde{\pi}_{LA}(\boldsymbol{\theta}, \boldsymbol{y}) = \frac{\pi(\widehat{\boldsymbol{U}}_{\boldsymbol{\theta}},\boldsymbol{\theta}, \boldsymbol{y})}{\widetilde{\pi}_G(\widehat{\boldsymbol{U}}_{\boldsymbol{\theta}}|\boldsymbol{\theta}, \boldsymbol{y})}, \label{equ:GaussianLaplaceApproxi}\]
where $\widehat{\boldsymbol{U}}_{\boldsymbol{\theta}} = \argmax_{\boldsymbol{U}} \log \pi(\boldsymbol{U}, \boldsymbol{\theta}, \boldsymbol{y})$, $\boldsymbol{H}_{\widehat{\boldsymbol{U}}_{\boldsymbol{\theta}}}(\boldsymbol{\theta}) = -\partial_{\boldsymbol{U}}^2 \log \pi(\widehat{\boldsymbol{U}}_{\boldsymbol{\theta}}, \boldsymbol{\theta}, \boldsymbol{y})$ and $d\mathcal{N}$ is the Gaussian density function.
Write the linear predictor as $\boldsymbol{\eta} = \boldsymbol{A}\boldsymbol{U}$, the negative hessian $\boldsymbol{H}_{\widehat{\boldsymbol{U}}_{\boldsymbol{\theta}}}(\boldsymbol{\theta})$ can be written explicitly as:
\[ \boldsymbol{H}_{\widehat{\boldsymbol{U}}_{\boldsymbol{\theta}}}(\boldsymbol{\theta}) = -\partial^2_{\boldsymbol{U}} \log \pi(\widehat{\boldsymbol{U}}_{\boldsymbol{\theta}}, \boldsymbol{\theta}, \boldsymbol{y}) =
\boldsymbol{Q}(\boldsymbol{\theta}_2) - \boldsymbol{A}^T \partial^2_{\boldsymbol{\eta}}\log \pi(\boldsymbol{y}|\widehat{\boldsymbol{\eta}}_{\boldsymbol{\theta}},\boldsymbol{\theta}_1) \boldsymbol{A}, \label{equ:hessianExplicit}
\]
where $\boldsymbol{Q}(\boldsymbol{\theta}_2) = \boldsymbol{\Sigma}^{-1}(\boldsymbol{\theta}_2)$ is the precision matrix and $\widehat{\boldsymbol{\eta}}_{\boldsymbol{\theta}} = \boldsymbol{A}\widehat{\boldsymbol{U}}_{\boldsymbol{\theta}}$.

The Laplace approximation is then normalized using the adaptive Gauss-Hermite quadrature (AGHQ) in \cite{aghqtheory}. 
Assume $\boldsymbol{\theta} \in \mathbb{R}^s$ and $\mathcal{Q}(K)$ denotes a set of Gauss-Hermite quadrature points with $K$ points per dimension of $\boldsymbol{\theta}$, and $\omega_K$ denotes the quadrature weight function.
Let $\widehat{\boldsymbol{\theta}}_{LA} = \argmax_{\boldsymbol{\theta}} \log \widetilde{\pi}_{LA}(\boldsymbol{\theta}, \boldsymbol{y})$ and ${\boldsymbol{H}}_{LA}(\widehat{\boldsymbol{\theta}}_{LA}) = - \partial_{\boldsymbol{\theta}}^2 \log \widetilde{\pi}_{LA}(\widehat{\boldsymbol{\theta}}_{LA}, \boldsymbol{y}) = \widehat{\boldsymbol{L}}_{LA} \widehat{\boldsymbol{L}}_{LA}^T$, where $\widehat{\boldsymbol{L}}_{LA}$ is the lower Cholesky matrix. 
The Laplace approximation in \cref{equ:GaussianLaplaceApproxi} can be then normalized as:
\[\widetilde{\pi}_{LA}(\boldsymbol{\theta}| \boldsymbol{y}) = \frac{\widetilde{\pi}_{LA}(\boldsymbol{\theta}, \boldsymbol{y})}{|\widehat{\boldsymbol{L}}_{LA}|\sum_{\boldsymbol{z} \in \mathcal{Q}(K)} \widetilde{\pi}_{LA}(\widehat{\boldsymbol{L}}_{LA}\boldsymbol{z} + \widehat{\boldsymbol{\theta}}_{LA}, \boldsymbol{y})w_k(\boldsymbol{z})}. \label{equ:AGHQ1}\]
Consequently, the posterior of the latent field can be approximated as:
\[\Tilde{\pi}(\boldsymbol{U}|\boldsymbol{y}) = |\widehat{\boldsymbol{L}}_{LA}|\sum_{\boldsymbol{z} \in \mathcal{Q}(K)} \widetilde{\pi}_G(\boldsymbol{U}|\widehat{\boldsymbol{L}}_{LA}\boldsymbol{z} + \widehat{\boldsymbol{\theta}}_{LA}, \boldsymbol{y})\widetilde{\pi}_{LA}(\widehat{\boldsymbol{L}}_{LA}\boldsymbol{z} + \widehat{\boldsymbol{\theta}}_{LA}| \boldsymbol{y})
w_k(\boldsymbol{z}), \label{equ:AGHQ2}\]
using the same set of quadrature rule $\mathcal{Q}(K)$ and $\{\omega_k\}_{k=1}^K$.
For more details of the implementation, refer to \cite{elgm}(Algorithm 1) and \cite{aghqtheory}.

\section{Bayesian Optimization Sequential Surrogate}\label{sec:method}

\subsection{Conditional LGMs}\label{subsec:conditionalELGM}

The methods in \cref{subsec:AGHQ} enables the efficient use of a wide class of LGMs. 
However, many commonly used hierarchical models can only be written as LGM when some conditioning parameters are given, which are called conditional LGM.

At a fixed value of the conditioning parameter $\boldsymbol{\alpha}$, the conditional LGM can be written as the following:
\begin{equation}\label{equ:conditionalELGM}
    \begin{aligned}
    y_i|\boldsymbol{\eta}_{\boldsymbol{\alpha}},\boldsymbol{\theta}_{\boldsymbol{\alpha}} &\overset{ind}{\sim} \pi(Y_i|\boldsymbol{\eta}_{\boldsymbol{\alpha}}, \boldsymbol{\theta}_{\boldsymbol{\alpha}}), \ \forall i \in [n]  \\
    \boldsymbol{\eta}_{\boldsymbol{\alpha}} = \boldsymbol{A}_{\boldsymbol{\alpha}} \boldsymbol{U}_{\boldsymbol{\alpha}}, \quad \boldsymbol{U}_{\boldsymbol{\alpha}}|\boldsymbol{\theta}_{\boldsymbol{\alpha}} &\sim \mathcal{N}\{\boldsymbol{0}, \boldsymbol{\Sigma}(\boldsymbol{\theta}_{\boldsymbol{\alpha}})\}, \quad
    \boldsymbol{\theta}_{\boldsymbol{\alpha}}  \sim \pi(\boldsymbol{\theta}_{\boldsymbol{\alpha}}),
    \end{aligned}
\end{equation}
which is a regular LGM in \cref{equ:ELGMdef} that can be inferred using the method in \cref{subsec:AGHQ} with the subscript of $\boldsymbol{\alpha}$ represents explicit dependence of parameters/quantities on $\boldsymbol{\alpha}$.
However, when $\boldsymbol{\alpha}$ is unknown with prior $\pi(\boldsymbol{\alpha})$, the model in \cref{equ:conditionalELGM} is no longer an LGM as the entire model structure in the latent field changes with $\boldsymbol{\alpha}$.
For example in a model with an unknown change point $\boldsymbol{\alpha}$, the design matrix $\boldsymbol{A}_{\boldsymbol{\alpha}}$ is different at each possible location of the change point.
As a result, the approximate inference method in \cref{subsec:AGHQ} can no longer be applied to such conditional LGMs.

To make efficient inference of conditional LGMs, 
existing approaches are either based on the Bayesian model averaging over a regular grid of $\boldsymbol{\alpha}$ \citep{bivand2014approximate} or the MCMC to obtain samples from $\pi(\boldsymbol{\alpha}|\boldsymbol{y})$ \citep{gomez2018markov}. 
Both of these require a huge number of evaluation of $\pi(\boldsymbol{\alpha}, \boldsymbol{y})$.
Since each evaluation of $\pi(\boldsymbol{\alpha}, \boldsymbol{y})$ requires fitting an LGM, both approaches become computationally infeasible when the LGM gets complicated and hence each evaluation gets costly.

To avoid the huge computational cost due to a large number of evaluations of $\pi(\boldsymbol{\alpha}, \boldsymbol{y})$ in the grid or MCMC methods, we propose the use of a Bayesian optimization (BO) to sequentially obtain design points that capture the majority of the mass of $\pi(\boldsymbol{\alpha} \mid \boldsymbol{y})$, and subsequently acquire a surrogate function for $\pi(\boldsymbol{\alpha} \mid \boldsymbol{y})$. Hence, we call our method the Bayesian Optimization Sequential Surrogate (BOSS).

\subsection{Design Points Construction via BOSS for Conditioning Parameter}
Consider the conditional LGM in Eq.~\ref{equ:conditionalELGM}, the quantities of interests are the following marginal posterior distributions
\[\label{eqn:U_marginal_cELGM}
\pi(\boldsymbol{U} \mid \boldsymbol{y}) = \int\pi(\boldsymbol{U}_{\boldsymbol{\alpha}}\mid \boldsymbol{y}, \boldsymbol{\alpha})\pi(\boldsymbol{\alpha} \mid \boldsymbol{y})d\boldsymbol{\alpha},
\]
\[\label{eqn:theta_marginal_cELGM}
\pi(\boldsymbol{\theta} \mid \boldsymbol{y}) = \int\pi(\boldsymbol{\theta}_{\boldsymbol{\alpha}}\mid \boldsymbol{y}, \boldsymbol{\alpha})\pi(\boldsymbol{\alpha} \mid \boldsymbol{y})d\boldsymbol{\alpha},
\]
and
\[\label{eqn:alpha_marginal_cELGM}
\pi(\boldsymbol{\alpha} \mid \boldsymbol{y}) = \frac{\pi(\boldsymbol{\alpha}, \boldsymbol{y})}{\int\pi(\boldsymbol{\alpha}, \boldsymbol{y})d\boldsymbol{\alpha}}.
\]
Here $(\boldsymbol{U}_{\boldsymbol{\alpha}}, \boldsymbol{\theta}_{\boldsymbol{\alpha}})$ is dropped to $(\boldsymbol{U}, \boldsymbol{\theta})$ due to marginalization over $\boldsymbol{\alpha}$. 
As mentioned, when $\boldsymbol{\alpha}$ is fixed, $\pi(\boldsymbol{U}_{\boldsymbol{\alpha}}\mid \boldsymbol{y}, \boldsymbol{\alpha})$ and $\pi(\boldsymbol{\theta}_{\boldsymbol{\alpha}}\mid \boldsymbol{y}, \boldsymbol{\alpha})$ can be efficiently approximated through methods described in Section \ref{subsec:AGHQ}. When $\boldsymbol{\alpha}$ is unknown, the grid search method essentially evaluates the integrals in Eq.~\ref{eqn:U_marginal_cELGM} and Eq.~\ref{eqn:theta_marginal_cELGM} through a Bayesian model averaging on a fine regular grid of $\boldsymbol{\alpha}$, while the MCMC method approximates them by drawing samples from the posterior distribution in Eq.~\ref{eqn:alpha_marginal_cELGM} through Markov chains.

Both grid search and MCMC methods aim to locate the region with the majority of the posterior mass under $\pi(\boldsymbol{\alpha} \mid \boldsymbol{y})$ through a large number of evaluations of $\pi(\boldsymbol{\alpha} \mid \boldsymbol{y})$.
For conditional LGMs, $\pi(\boldsymbol{\alpha} \mid \boldsymbol{y})$ is an intractable function and its evaluation requires fitting \cref{equ:conditionalELGM} through nested approximations described in \cref{subsec:AGHQ}. 
As a result, the large number of evaluations make their computation for complex models infeasible.
Designed for global optimization of intractable objective functions, Bayesian Optimization (BO) can locate the high posterior mass region under $\pi(\boldsymbol{\alpha} \mid \boldsymbol{y})$ with a small number of evaluations of $\pi(\boldsymbol{\alpha} \mid \boldsymbol{y})$.
Specifically, BO constructs a sequence of design points that accurately depict the behaviour of the objective function near its mode.
When the objective function is $\pi(\boldsymbol{\alpha} \mid \boldsymbol{y})$, the design points then capture the majority of its posterior mass.

In the context of conditional LGMs, notice that
\[\label{eqn:alpha_posterior}
\log\pi(\boldsymbol{\alpha} \mid \boldsymbol{y}) \propto \log\pi(\boldsymbol{y}, \bm{\alpha}) = \log\pi(\boldsymbol{y} \mid \bm{\alpha}) + \log\pi(\bm{\alpha}).
\]
$\log\pi(\boldsymbol{y} \mid \bm{\alpha})$ is effectively the log-marginal likelihood of the model in Eq.~\ref{equ:conditionalELGM} when $\boldsymbol{\alpha}$ is fixed, and it can be accurately approximated using methods in \cref{subsec:AGHQ}:
\[\label{eqn:mlik_cELGM}
\Tilde{\pi}(\boldsymbol{y} \mid \boldsymbol{\alpha}) = \int\widetilde{\pi}_{LA}(\boldsymbol{\theta}_{\boldsymbol{\alpha}}, \boldsymbol{y} \mid \boldsymbol{\alpha})d\boldsymbol{\theta}_{\boldsymbol{\alpha}},
\]
where the integration could be done numerically such as in \cref{equ:AGHQ1}.
Here $\widetilde{\pi}_{LA}(\boldsymbol{\theta}_{\boldsymbol{\alpha}}, \boldsymbol{y} \mid \boldsymbol{\alpha})$ is defined as in Eq.~\ref{equ:GaussianLaplaceApproxi} but with explicit dependence on $\boldsymbol{\alpha}$ from conditioning. 

To apply BO, the objective function needs to be defined on a compact search space $\Omega \in \mathbb{R}^d$ where $d$ is the dimension of $\bm{\alpha}$. 
Let $\Omega=\prod_{i=1}^d[l_i, u_i] = [l_1, u_1]\times\dots\times[l_d, u_d] \subset \mathbb{R}^d$ be the essential support of $\pi(\bm{\alpha} \mid \bm{y})$ such that $\int_\Omega \pi(\bm{\alpha} \mid \bm{y}) d\bm{\alpha} \approx 1$, and define
\[f(\boldsymbol{\alpha}) = \log\Tilde{\pi}(\boldsymbol{y} \mid \bm{\alpha}) + \log\pi(\bm{\alpha}), \ \boldsymbol{\alpha} \in \Omega,
\]
as the objective function.
We then apply BO to optimize $f(\boldsymbol{\alpha})$ with $B$ iterations by sequentially selecting the design points $\mathcal{X}_B = \{(\bm{\alpha}^{(i)},f(\bm{\alpha}^{(i)})): i \in [B]\}$. 
The objective function $f$ is assigned a zero-mean Gaussian process (GP) with a specified covariance function $\mathcal{C}$, which we set to the square exponential kernel as default. 
We start by selecting an initial point $\bm{\alpha}^{(1)} \in \Omega$ and evaluate $f(\bm{\alpha}^{(1)})$. 
At a given iteration $t$, we have a set of design points $\mathcal{X}_t = \{(\bm{\alpha}^{(i)}, f(\bm{\alpha}^{(i)})): i \in [t]\}$, and we update the distribution of $f$ by conditioning on $\mathcal{X}_t$. 
We then construct an acquisition function (AF) based on $f\mid\mathcal{X}_t$, and optimize the AF to select the next design point $\bm{\alpha}^{(t+1)}$. 
We repeat this until we obtain $B$ number of design points, where $B$ is pre-specified integer based on the amount of available computational resource and the complexity of the problem.
Algorithm \ref{algo: BOSS_cELGM} summarises the procedure of our BOSS algorithm to obtain the design points $\mathcal{X}_B$.

\begin{algorithm}
\caption{Bayesian Optimization Sequential Surrogate for Conditional LGM}
\begin{algorithmic}[1]\label{algo: BOSS_cELGM} 
  \REQUIRE Number of BO iterations $B$; Search space $\Omega=\prod_{i=1}^d[l_i, u_i]$ for $\boldsymbol{\alpha}$; Conditional LGM $\mathcal{M}(\boldsymbol{\alpha})$ as in Eq.~\ref{equ:conditionalELGM}; Initial point $\bm{\alpha}^{(1)} \in \Omega$; Assign a GP with covariance function $\mathcal{C}$ for $f$. 
  \FOR{$t = 1$ to $B$}
   \STATE Evaluate $f(\bm{\alpha}^{(t)})$ via Eq.~\ref{eqn:alpha_posterior} and \ref{eqn:mlik_cELGM} by fitting $\mathcal{M}(\bm{\alpha}^{(t)})$.
   \STATE Set $\mathcal{X}_t = \{(\bm{\alpha}^{(i)}, f(\bm{\alpha}^{(i)})): i \in [t]\}$.
   \STATE Obtain the conditional distribution $f \mid \mathcal{X}_t$.
   \STATE Construct AF $\mathcal{G}_t(\bm{\alpha})$ based on $f \mid \mathcal{X}_t$ and set
   \[
   \bm{\alpha}^{(t+1)} = \argmax_{\bm{\alpha} \in \Omega}\mathcal{G}_t(\bm{\alpha}), 
   \]
    \ENDFOR
    \STATE \textbf{Output:} Sequential design points $\mathcal{X}_B = \{(\bm{\alpha}^{(i)},f(\bm{\alpha}^{(i)})): i \in [B]\}$.
\end{algorithmic}
\end{algorithm}
In \cref{algo: BOSS_cELGM}, the design points are selected based on the optimization of the AF $\mathcal{G}_t(\boldsymbol{\alpha})$.
Motivated by \cite{srinivas2012information}, we consider the following upper confidence bound (UCB) as the default AF:
\begin{equation}\label{equ:UCB}
    \begin{aligned}
        \mathcal{G}_t(\boldsymbol{\alpha}) = m_{t}(\boldsymbol{\alpha}) + \gamma_t^{1/2}\mathcal{C}_{t}(\boldsymbol{\alpha},\boldsymbol{\alpha}),
    \end{aligned}
\end{equation}
where $m_{t}$ and $\mathcal{C}_{t}$ are respectively the mean function and the covariance function of $f \mid \mathcal{X}_t$, and $\gamma_t = 2\log(t^2\pi^2/6\delta)$ for a small, pre-specified $\delta \in (0,1)$. 
Note that $\mathcal{G}_t(\boldsymbol{\alpha})$ is large either when the prediction $m_{t}(\boldsymbol{\alpha})$ or the uncertainty $\mathcal{C}_{t}(\boldsymbol{\alpha},\boldsymbol{\alpha})$ is large.
Therefore, $\mathcal{G}_t(\boldsymbol{\alpha})$ balances the exploration (i.e. minimizing uncertainty) and exploitation (i.e. maximizing the prediction) of $f(\bm{\alpha})$ in the search space $\Omega$. 
There are various other types of AF such as Thompson sampling and expected improvement that could be applied in \cref{algo: BOSS_cELGM}; see \cite{shahriari2015taking} for more details.
For the simplicity of the presentation, we stick to the UCB in \cref{equ:UCB} as the AF in the rest of this paper.

The goal of Algorithm \ref{algo: BOSS_cELGM} is to obtain the design points $\mathcal{X}_B$ with a small number ($B$) of evaluations of $f(\bm{\alpha})$, which accurately describes the region containing majority of the posterior mass under $\pi(\bm{\alpha} \mid \bm{y})$. 
After obtaining $\mathcal{X}_B$, we construct a surrogate function $f_{\texttt{BO}}(\boldsymbol{\alpha})$ through a smooth interpolation of $\mathcal{X}_B$.
A natural candidate of $f_{\texttt{BO}}(\boldsymbol{\alpha})$ is the mean function $m_B(\boldsymbol{\alpha})$ of $f\mid \mathcal{X}_B$ which has been constructed during the BO in \cref{algo: BOSS_cELGM}, but other interpolation techniques such as smoothing splines \citep{greensplinebook} can also be used.
In effect, $f_{\texttt{BO}}(\boldsymbol{\alpha})$ approximates the unnormalized log-posterior of $\bm{\alpha}$, which upon normalization will provide us with an accurate surrogate for the true posterior $\pi(\bm{\alpha} \mid \bm{y})$.

For multi-modal $\pi(\bm{\alpha}\mid \bm{y})$, Algorithm \ref{algo: BOSS_cELGM} could start with multiple initial values to ensure sufficient exploration of local modes. 
We show in Section \ref{sec:simulation} through simulations that the proposed BOSS algorithm is still accurate for such distributions provided enough BO iterations $B$ are used in \cref{algo: BOSS_cELGM}.

As our default implementation of BOSS utilizes a square exponential covariance function for $f(\bm{\alpha})$, it requires choosing two hyper-parameters: the scale parameter and the variance parameter \citep{rasmussen2003gaussian}. 
Starting with their pre-specified initial values, we update these hyper-parameters to their maximum likelihood estimates computed from the current design points, after a predetermined number of evaluations of $f(\bm{\alpha})$ as outlined in \cref{algo: BOSS_cELGM}.
The details of this procedure are described in the supplementary material.

\subsection{Normalization of the Sequential Surrogate}

Unlike the original function $f(\bm{\alpha})$, the evaluation of the surrogate function $f_{\texttt{BO}}(\bm{\alpha})$ obtained from BO has no computational cost, as $f_{\texttt{BO}}(\bm{\alpha})$ is a tractable function with an explicit form.
Thus, we can easily normalize $f_{\texttt{BO}}(\boldsymbol{\alpha})$ to obtain the surrogate posterior density of $\bm{\alpha}$:
\[\label{eqn:surrogate_posterior_alpha}
\pi_{\texttt{BO}}(\boldsymbol{\alpha} \mid \boldsymbol{y}) = \frac{\exp\{f_{\texttt{BO}}(\boldsymbol{\alpha})\}}{\int \exp\{f_{\texttt{BO}}(\boldsymbol{\alpha})\} d\bm{\alpha}}.
\]
There are various methods available for the integration in Eq.~\ref{eqn:surrogate_posterior_alpha}. We here illustrate possible approaches.

\subsubsection{Numerical Integration}\label{subsec:numerical integration alpha}

Firstly, one can directly normalize $\exp\{f_{\texttt{BO}}(\boldsymbol{\alpha})\}$ to obtain $\Tilde{\pi}_{\texttt{BO}}(\boldsymbol{\alpha} \mid \boldsymbol{y})$ through numerical integration:
\[
\Tilde{\pi}_{\texttt{BO}}(\boldsymbol{\alpha} \mid \boldsymbol{y}) = \frac{\exp\{f_{\texttt{BO}}(\boldsymbol{\alpha})\}}{\sum_{k=1}^K\exp\{f_{\texttt{BO}}(\boldsymbol{\alpha}_k)\}w(\boldsymbol{\alpha}_k)},
\]
where $\mathcal{K} = \{\bm{\alpha}_k\}_{k=1}^K$ is the set of integration points while $\{w(\boldsymbol{\alpha}_k)\}_{k=1}^K$ are the corresponding integration weights.

The simplest formulation of such approach is to consider $\mathcal{K} = \{\bm{\alpha}_k\}_{i=k}^K$ as a fine grid discretising the compact support $\Omega$. $\{w(\boldsymbol{\alpha}_k)\}_{k=1}^K$ are thus the volume of each grid.
If the posterior of $\bm{\alpha}$ is uni-modal, we can also employ AGHQ to obtain $\Tilde{\pi}_{\texttt{BO}}(\boldsymbol{\alpha} \mid \boldsymbol{y})$. Since integration by Gauss-Hermite quadrature operates on the entire real space, we require transformation of $\Omega$ to $\mathbb{R}^d$. Let $h_i: \mathbb{R} \to [l_i, u_i], i \in [d]$ be a strictly increasing differentiable function. Write $\bm{\alpha} = (\alpha_1, \dots, \alpha_d) \in \Omega$, and set $\bm{\alpha}' = (\alpha_1', \dots, \alpha_d') = (h_1^{-1}(\alpha_1), \dots, h_d^{-1}(\alpha_d)) \equiv \mathbf{h}^{-1}(\bm{\alpha}) \in \mathbb{R}^d$ with $\mathbf{h}(\bm{\alpha}') = \bm{\alpha}$. Suppose $\mathcal{Q}(K)$ is a set of Gauss-Hermite quadrature points with $K$ points per dimension of $\bm{\alpha}'$.
Denote $f^t_{\texttt{BO}}(\boldsymbol{\alpha}') = f_{\texttt{BO}}\{\mathbf{h}(\boldsymbol{\alpha}')\} + \log J_{\mathbf{h}}(\boldsymbol{\alpha}')$ where $J_{\mathbf{h}}(\boldsymbol{\alpha}')$ is the Jacobian of $\mathbf{h}(\bm{\alpha}')$.
Let $\widehat{\bm{\alpha}}_{\texttt{BO}}' = \argmax_{\boldsymbol{\alpha}'} f^t_{\texttt{BO}}(\boldsymbol{\alpha}')$ and ${\boldsymbol{H}}_{\texttt{BO}}(\widehat{\boldsymbol{\alpha}}_{\texttt{BO}}') = - \partial_{\boldsymbol{\alpha}'}^2 f^t_{\texttt{BO}}(\widehat{\boldsymbol{\alpha}}_{\texttt{BO}}') = \widehat{\boldsymbol{L}}_{\texttt{BO}} \widehat{\boldsymbol{L}}_{\texttt{BO}}^T$, where $\widehat{\boldsymbol{L}}_{\texttt{BO}}$ is the lower Cholesky matrix. 
$\Tilde{\pi}_{\texttt{BO}}(\boldsymbol{\alpha} \mid \boldsymbol{y})$ can be computed as:
\[\tilde{\pi}_{\texttt{BO}}(\boldsymbol{\alpha}| \boldsymbol{y}) = \frac{\exp(f_{\texttt{BO}}(\boldsymbol{\alpha}))}{|\widehat{\boldsymbol{L}}_{\texttt{BO}}|\sum_{\boldsymbol{z} \in \mathcal{Q}(K)} \exp\{f^t_{\texttt{BO}}(\widehat{\boldsymbol{L}}_{\texttt{BO}}\boldsymbol{z} + \widehat{\bm{\alpha}}_{\texttt{BO}}')\}w(\boldsymbol{z})}. \label{equ:AGHQ_alpha}\]

\subsubsection{MCMC}

Another method to obtain the surrogate posterior $\pi_{\texttt{BO}}(\boldsymbol{\alpha} \mid \boldsymbol{y})$ is to directly sample it via (approximate) MCMC based on $f_{\texttt{BO}}(\bm{\alpha})$. 
Unlike the MCMC that evaluates $f(\bm{\alpha})$, the approximate MCMC evaluates the surrogate $f_{\texttt{BO}} (\bm{\alpha})$ and hence is computationally efficient. Moreover, if $f_{\texttt{BO}}(\bm{\alpha})$ is smooth, one can use gradient-based MCMC such as Hamiltonian Monte-Carlo \citep{Brooks_2011, betancourt2018conceptual} to obtain posterior samples for $\bm{\alpha}$ since the surrogate $f_{\texttt{BO}}(\bm{\alpha})$ can be explicitly differentiated.



\subsection{Inference through BOSS}\label{subsec:BOSS_latent}

To conduct inference for the latent field $\bm{U}$ and the hyper-parameter $\bm{\theta}$ for conditional LGMs, we consider the approximations to the integrals in Eq.~\ref{eqn:U_marginal_cELGM} and \ref{eqn:theta_marginal_cELGM} through AGHQ in a similar fashion as in Section \ref{subsec:numerical integration alpha}. 
Following the same setup and notation in Section \ref{subsec:numerical integration alpha}, we can obtain approximate posterior distribution for $\bm{U}$ and $\bm{\theta}$ as
\[
\widetilde{\pi}_{\texttt{BO}}(\bm{U}\mid \bm{y}) = |\widehat{\boldsymbol{L}}_{\texttt{BO}}|\sum_{\boldsymbol{z} \in \mathcal{Q}(K)} \widetilde{\pi}\left\{\boldsymbol{U}\mid\mathbf{h}(\widehat{\boldsymbol{L}}_{\texttt{BO}}\boldsymbol{z} + \widehat{\boldsymbol{\alpha}}_{\texttt{BO}}'), \boldsymbol{y}\right\}\tilde{\pi}_{\texttt{BO}}\left\{\mathbf{h}(\widehat{\boldsymbol{L}}_{\texttt{BO}}\boldsymbol{z} + \widehat{\boldsymbol{\alpha}}_{\texttt{BO}}')\mid \boldsymbol{y}\right\}
w(\boldsymbol{z}),
\]
and
\[
\widetilde{\pi}_{\texttt{BO}}(\bm{\theta}\mid \bm{y}) = |\widehat{\boldsymbol{L}}_{\texttt{BO}}|\sum_{\boldsymbol{z} \in \mathcal{Q}(K)} \widetilde{\pi}\left\{\boldsymbol{\theta}\mid\mathbf{h}(\widehat{\boldsymbol{L}}_{\texttt{BO}}\boldsymbol{z} + \widehat{\boldsymbol{\alpha}}_{\texttt{BO}}'), \boldsymbol{y}\right\}\tilde{\pi}_{\texttt{BO}}\left\{\mathbf{h}(\widehat{\boldsymbol{L}}_{\texttt{BO}}\boldsymbol{z} + \widehat{\boldsymbol{\alpha}}_{\texttt{BO}}')\mid \boldsymbol{y}\right\}
w(\boldsymbol{z}).
\]
$ \widetilde{\pi}\left\{\boldsymbol{U}\mid\mathbf{h}(\widehat{\boldsymbol{L}}_{\texttt{BO}}\boldsymbol{z} + \widehat{\boldsymbol{\alpha}}_{\texttt{BO}}'), \boldsymbol{y}\right\}$ and $ \widetilde{\pi}\left\{\boldsymbol{\theta}\mid\mathbf{h}(\widehat{\boldsymbol{L}}_{\texttt{BO}}\boldsymbol{z} + \widehat{\boldsymbol{\alpha}}_{\texttt{BO}}'), \boldsymbol{y}\right\}$ are obtained by fitting another $K^d$ number of LGMs with $\bm{\alpha}$ being fixed to the selected quadrature points.

Combining the above procedure with Algorithm \ref{algo: BOSS_cELGM}, the total number of evaluations of $f(\boldsymbol{\alpha})$ is then $B + K^d$. 
Compared to grid and MCMC methods, our approach requires orders of magnitude fewer number of model fitting.
As we illustrate in \cref{sec:simulation}, having $B \lesssim 100$ and $K \lesssim 4$ is well-beyond sufficient to obtain accurate inference in typical applications. 
On the other hand, for grid-based method \citep{bivand2014approximate}, typically at least $100$ grid points in each dimension of $\bm{\alpha}$ are needed to make accurate inference, resulting in a total of $N = 100^d$ model evaluations. 
For the MCMC-based method \citep{gomez2018markov}, it would at least take thousands if not tens of thousands iterations to achieve the same accuracy.
Furthermore, our BOSS algorithm also has significantly less memory requirements since we only need to store $K^d$ number of fitted LGMs. In comparison, all LGMs fitted for the grid search or MCMC method have to be stored in order to make inference for $\bm{U}$ and $\bm{\theta}$. For complex scenarios, memory needed for storing thousands of such LGMs will be impossible to achieve.

\section{Simulation}\label{sec:simulation}

\subsection{Simulation 1: Explore Complex Posterior Functions}\label{subsec:simulation1}

In this section, we assess the accuracy of the proposed BOSS algorithm for approximating (un-normalized) posterior functions with different complexity. 
We consider a conditioning parameter $\alpha$ with support $\Omega = [0,10]$, and assume its unnormalized log posteriors are respectively defined to $f(\alpha) = \alpha \sin(\alpha)$, $\log(\alpha+1)\sin(2\alpha) - \alpha \cos(2\alpha)$,
and $\log(\alpha + 1) (\sin(4\alpha) + \cos(2\alpha))$ for simple, medium, and hard settings.
In the simple setting, the log posterior has two local modes and its corresponding posterior is close to uni-modal. 
In the medium scenario, the log posterior has three local modes, with a corresponding posterior that is close to bi-modal. 
In the hard scenario, the log posterior has seven local modes and the posterior is close to tri-modal.
The proposed BOSS algorithm is then applied with different number of BO iterations $B$, as described in the earlier sections. 
The algorithm uses three initial values equally placed between 0 to 10.

\begin{figure}
          \centering
          \subfigure[Simple Case: $B = 10$]{
      \includegraphics[width=0.31\textwidth]{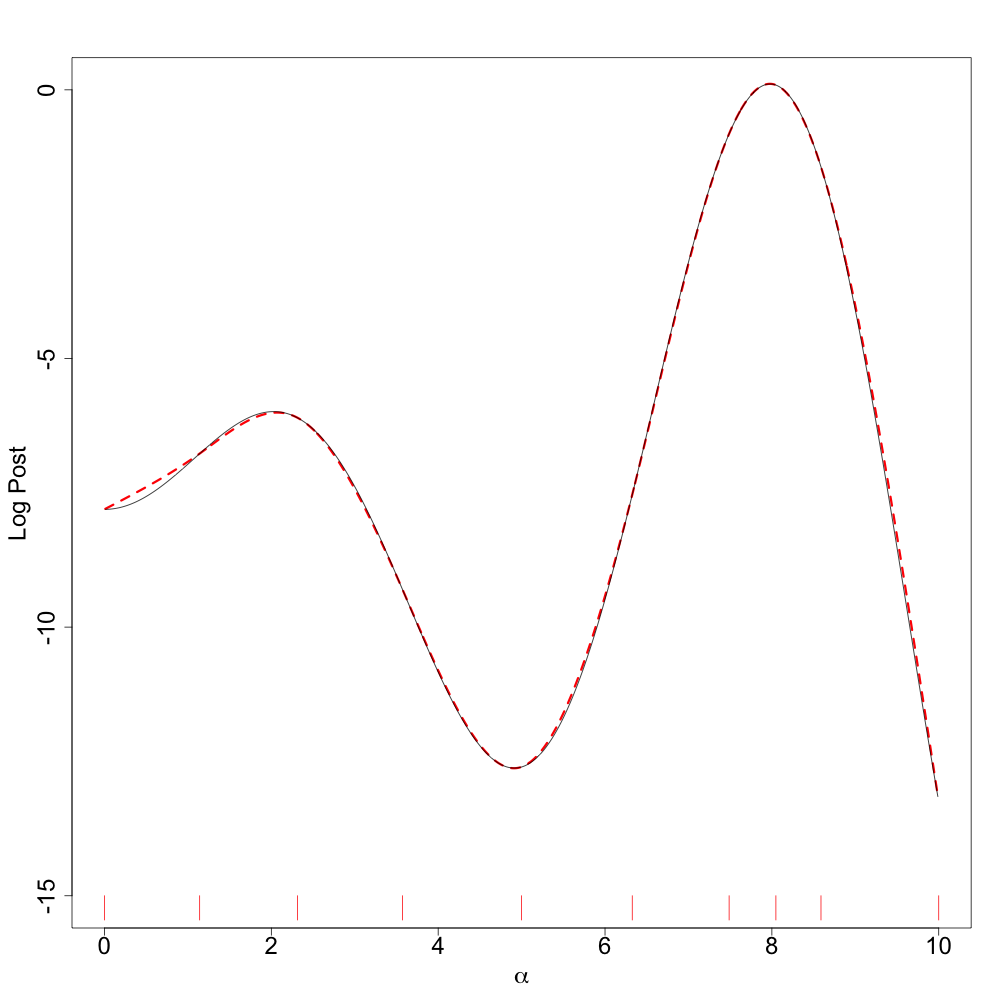}
    }
          \subfigure[Simple Case: $B = 30$]{
      \includegraphics[width=0.31\textwidth]{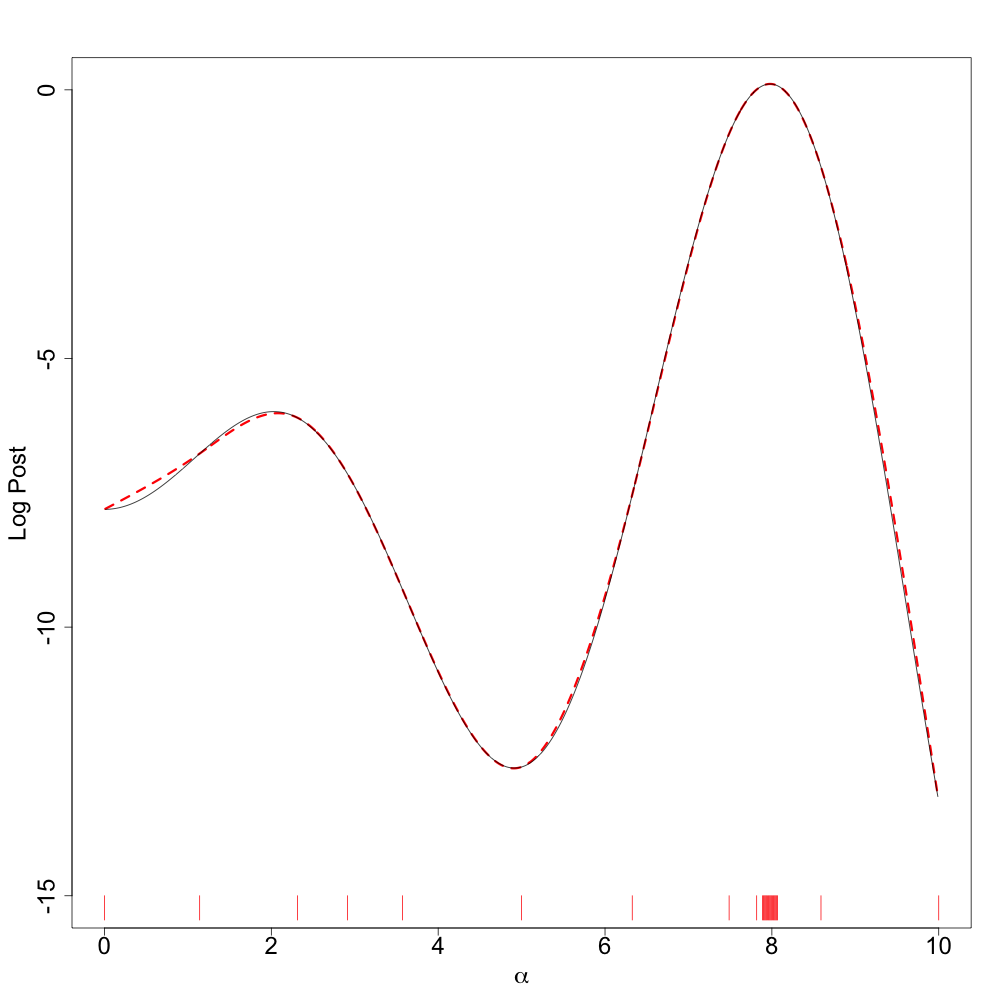}
    }
           \subfigure[Simple Case: $B = 80$]{
      \includegraphics[width=0.31\textwidth]{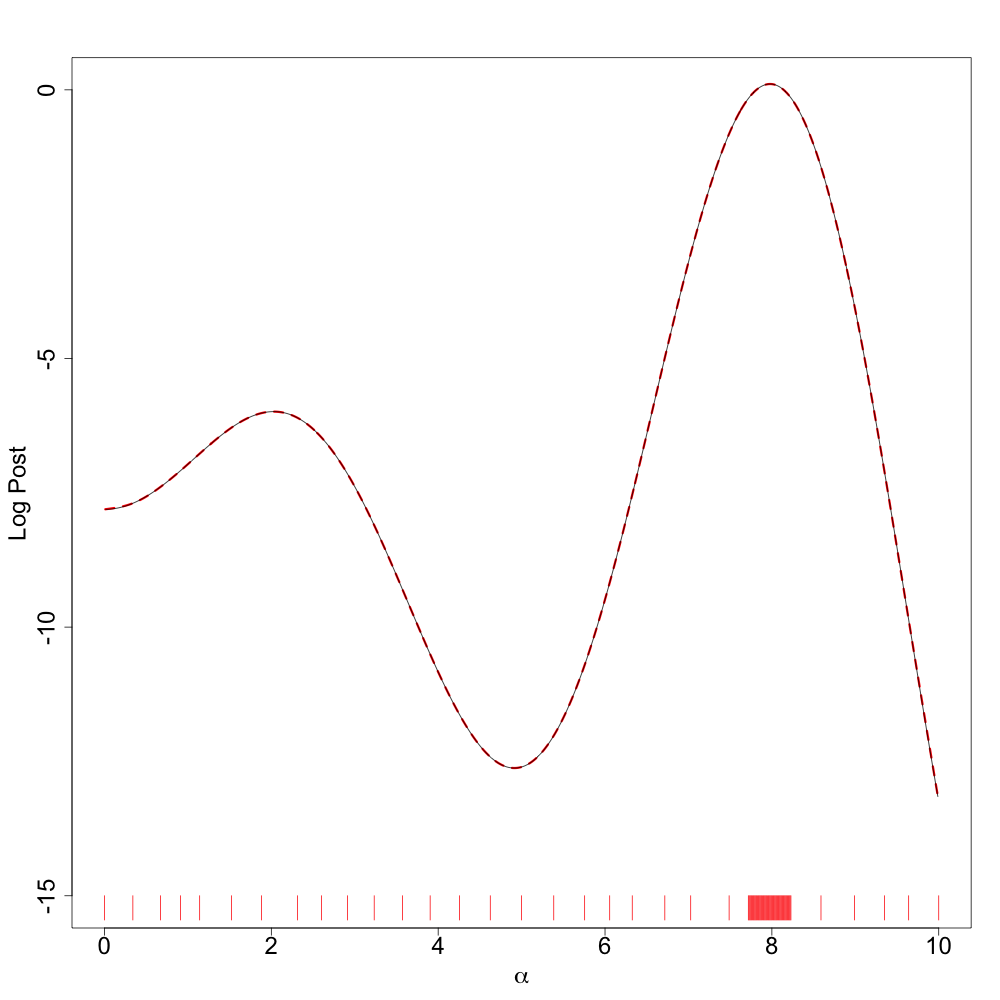}
    }
          \subfigure[Medium Case: $B = 10$]{
      \includegraphics[width=0.31\textwidth]{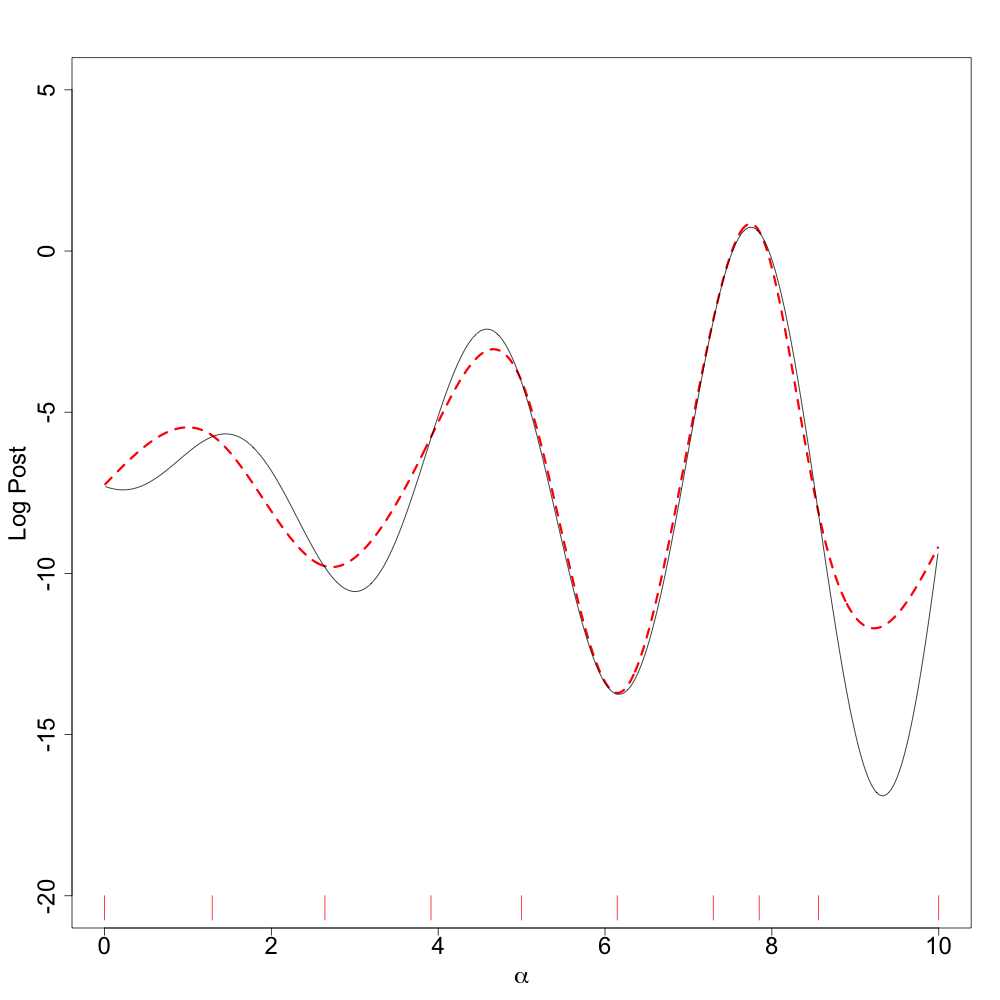}
    }
          \subfigure[Medium Case: $B = 30$]{
      \includegraphics[width=0.31\textwidth]{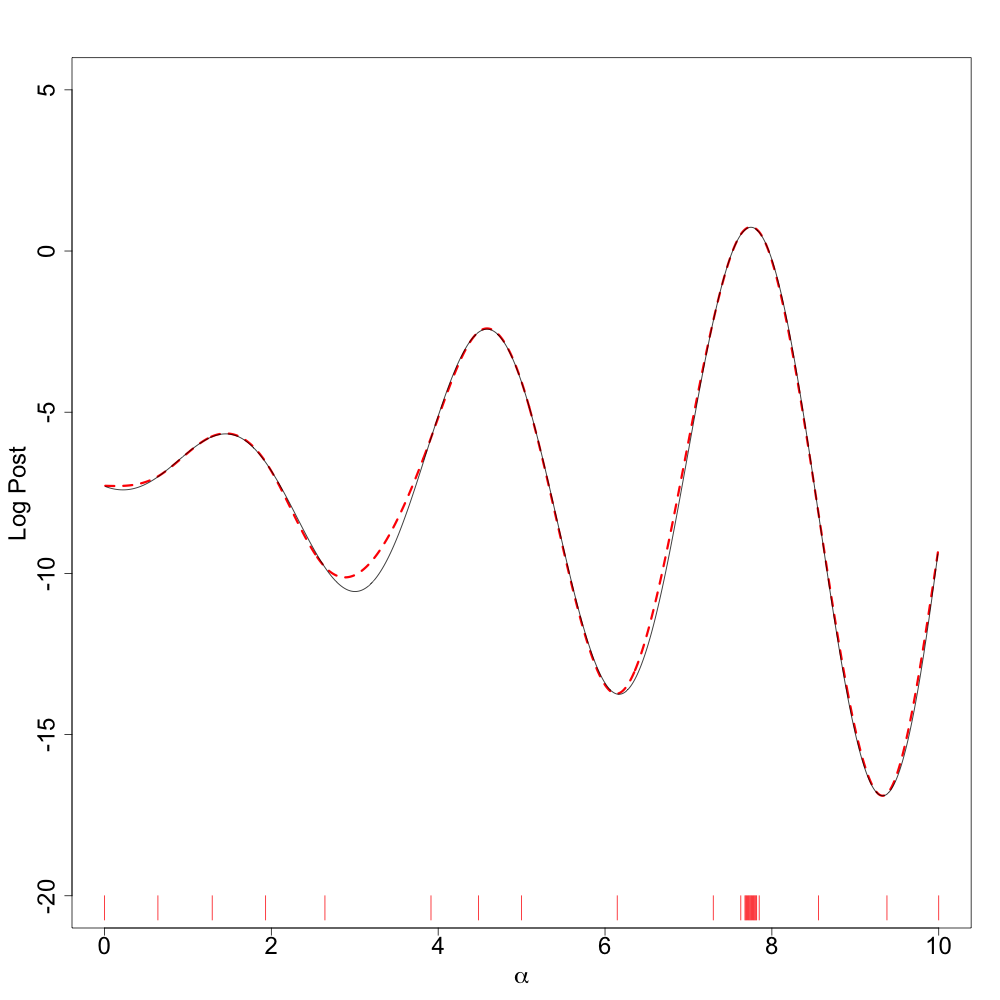}
    }
           \subfigure[Medium Case: $B = 80$]{
      \includegraphics[width=0.31\textwidth]{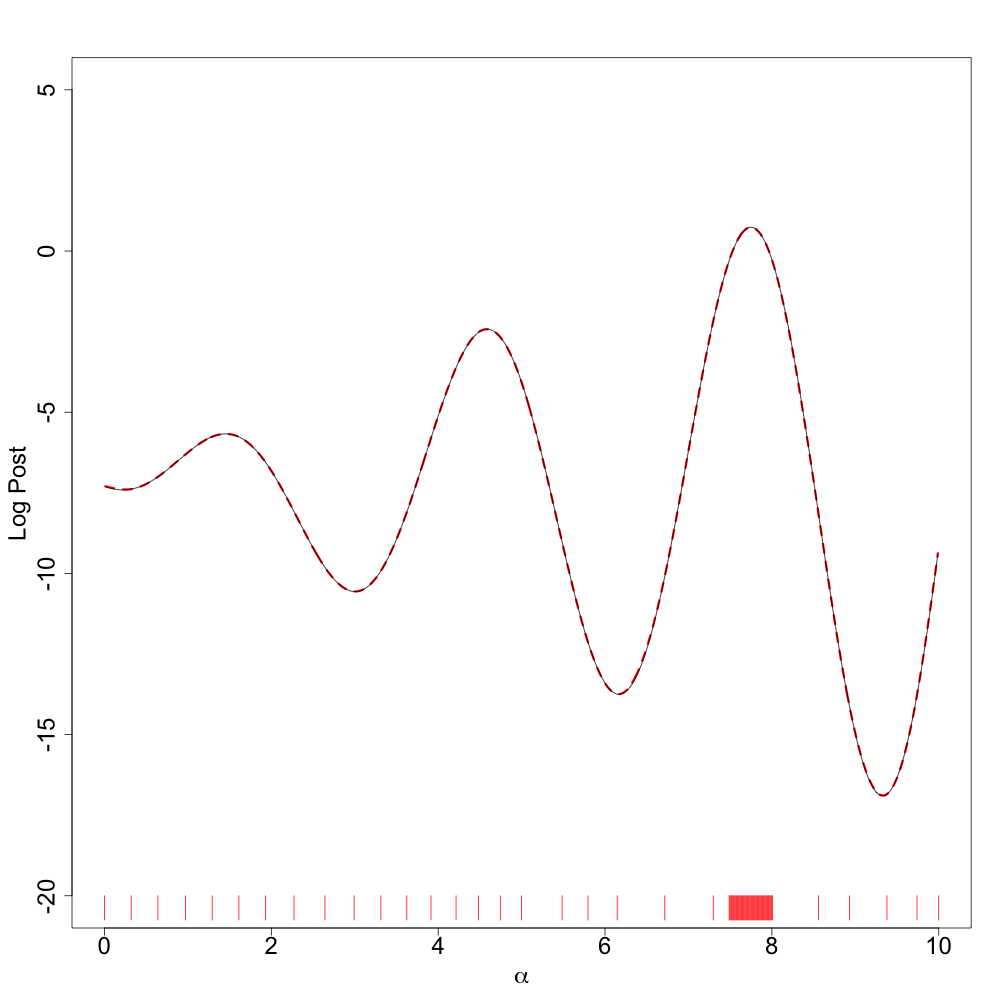}
    }
              \subfigure[Hard Case: $B = 10$]{
      \includegraphics[width=0.31\textwidth]{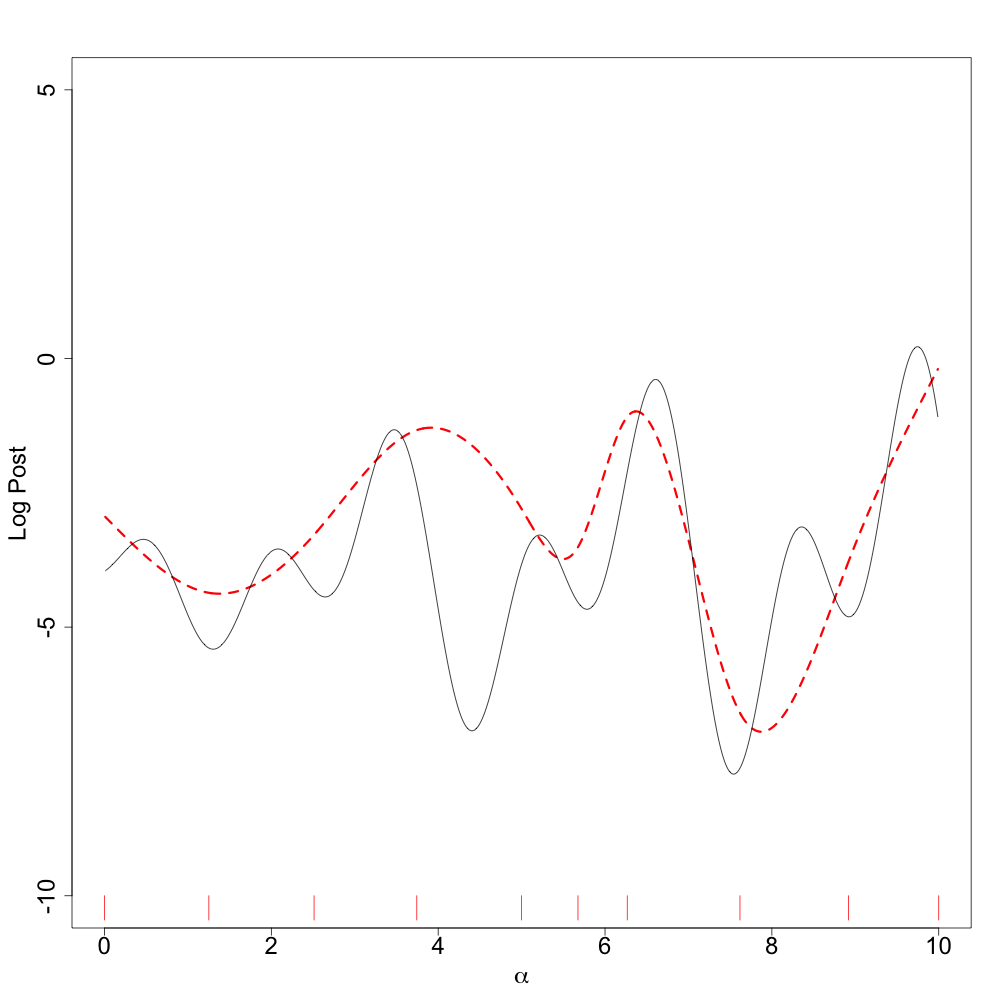}
    }
          \subfigure[Hard Case: $B = 30$]{
      \includegraphics[width=0.31\textwidth]{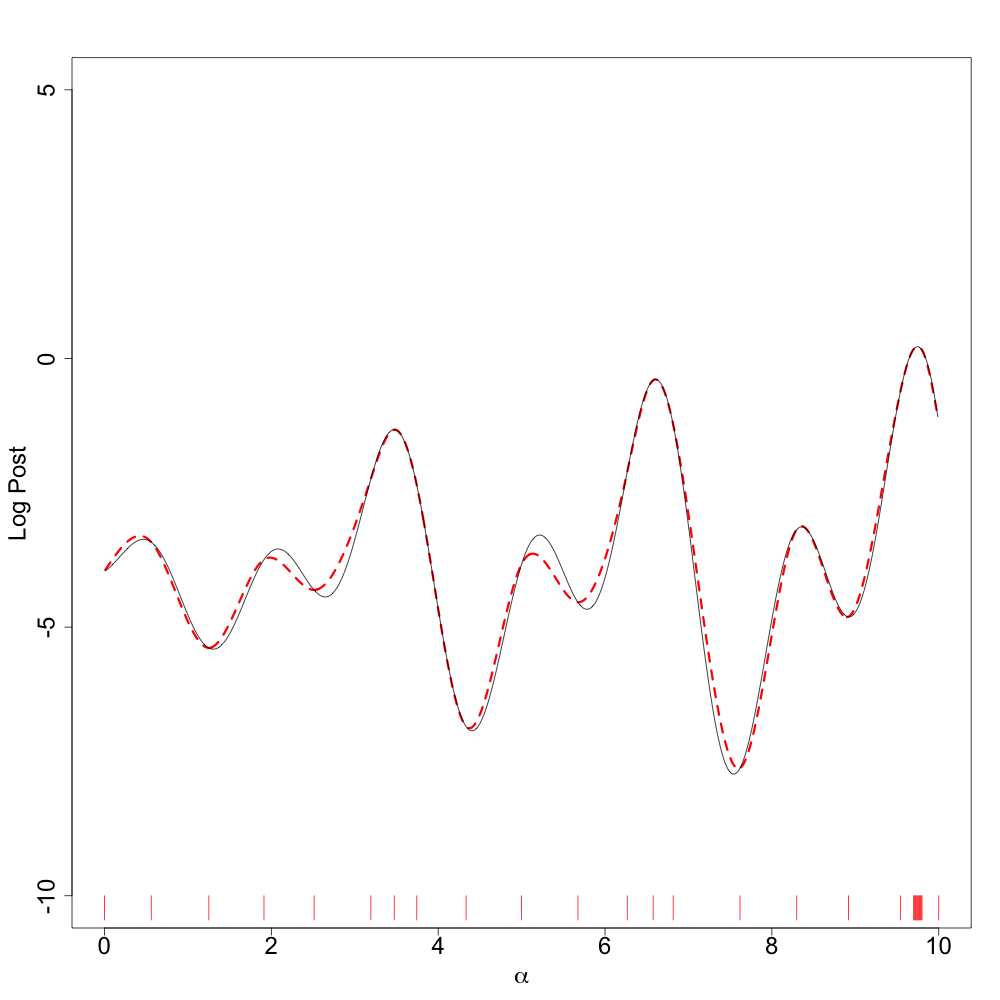}
    }
           \subfigure[Hard Case: $B = 80$]{
      \includegraphics[width=0.31\textwidth]{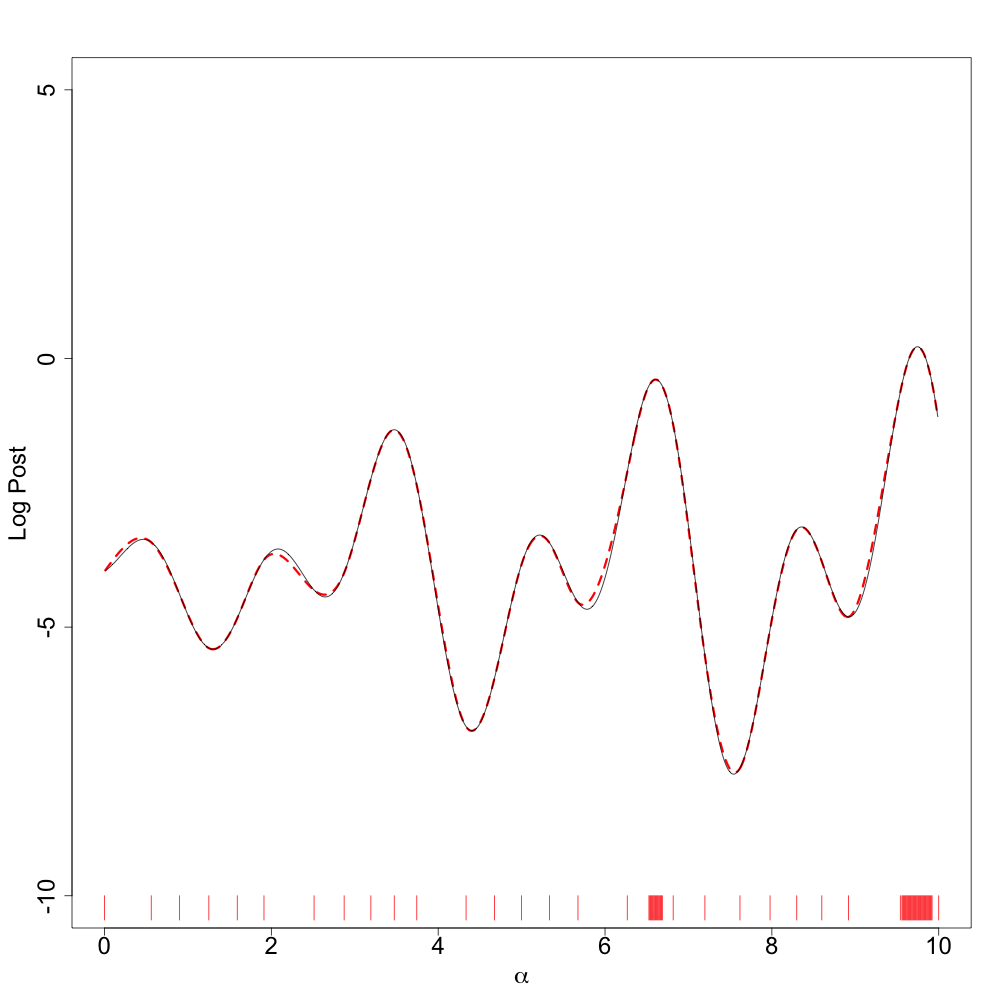}
    }
    
    \caption{Log posterior distribution of $\alpha$ in \cref{subsec:simulation1}. The red dashed lines denote the approximated log posteriors obtained from the BOSS algorithm using different numbers of BO iterations $B$. The black solid line denotes the true log posteriors. The design points of the BOSS algorithm are shown in the red vertical lines on the x-axis.}
    \label{fig:sim1logPost}
\end{figure}

\begin{figure}
          \centering
          \subfigure[Simple Case: $B = 10$]{
      \includegraphics[width=0.31\textwidth]{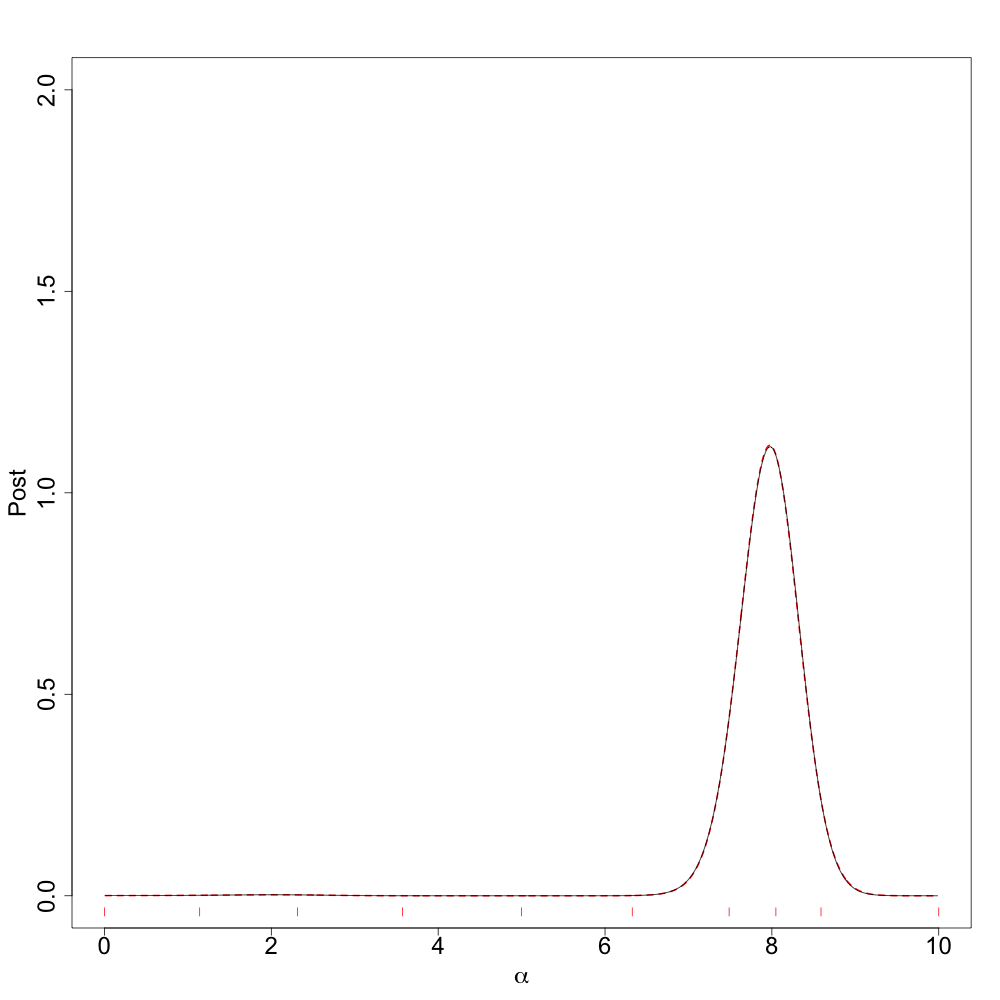}
    }
          \subfigure[Simple Case: $B = 30$]{
      \includegraphics[width=0.31\textwidth]{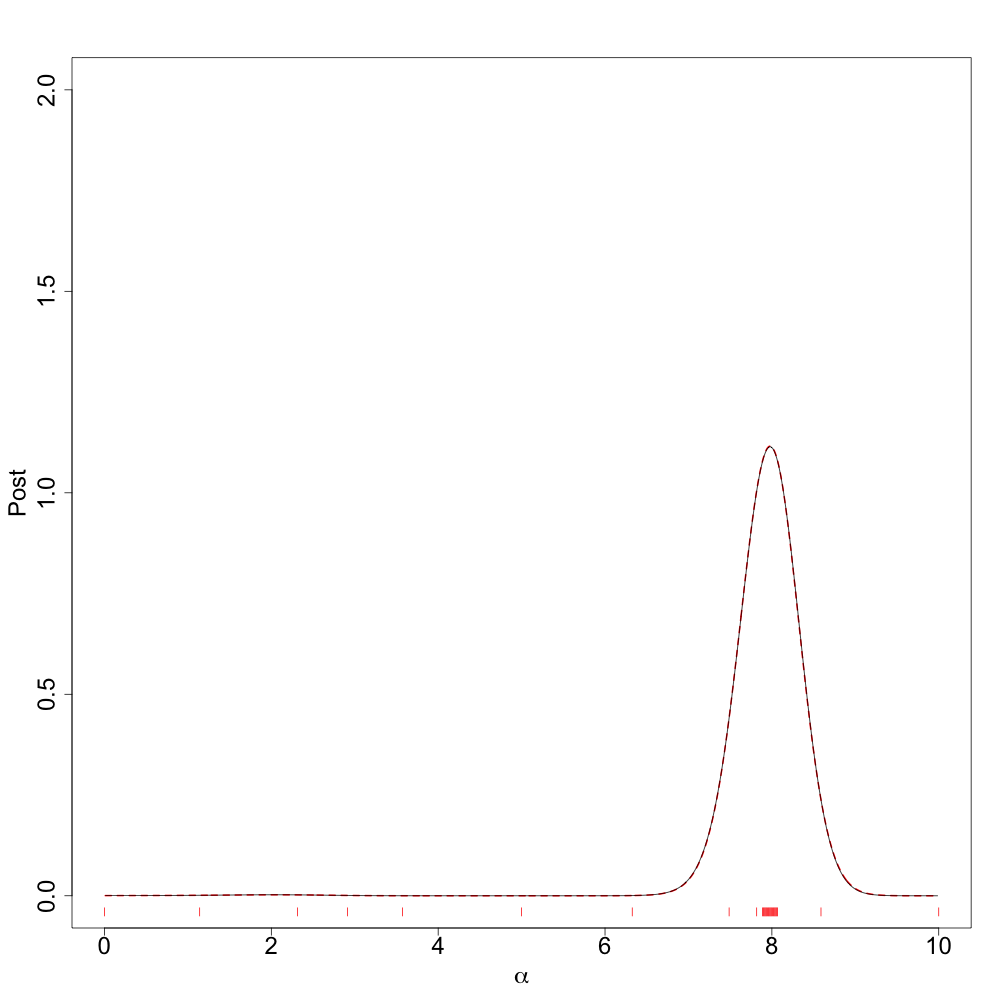}
    }
           \subfigure[Simple Case: $B = 80$]{
      \includegraphics[width=0.31\textwidth]{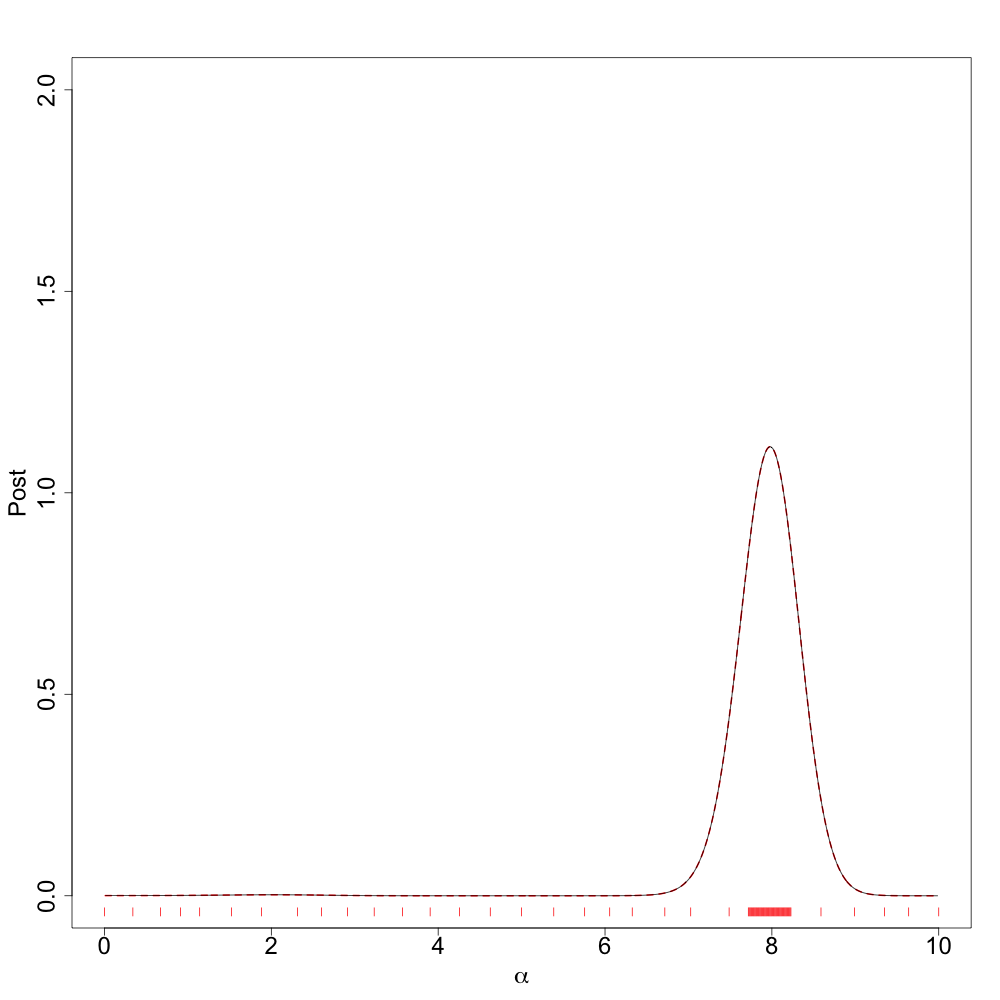}
    }
          \subfigure[Medium Case: $B = 10$]{
      \includegraphics[width=0.31\textwidth]{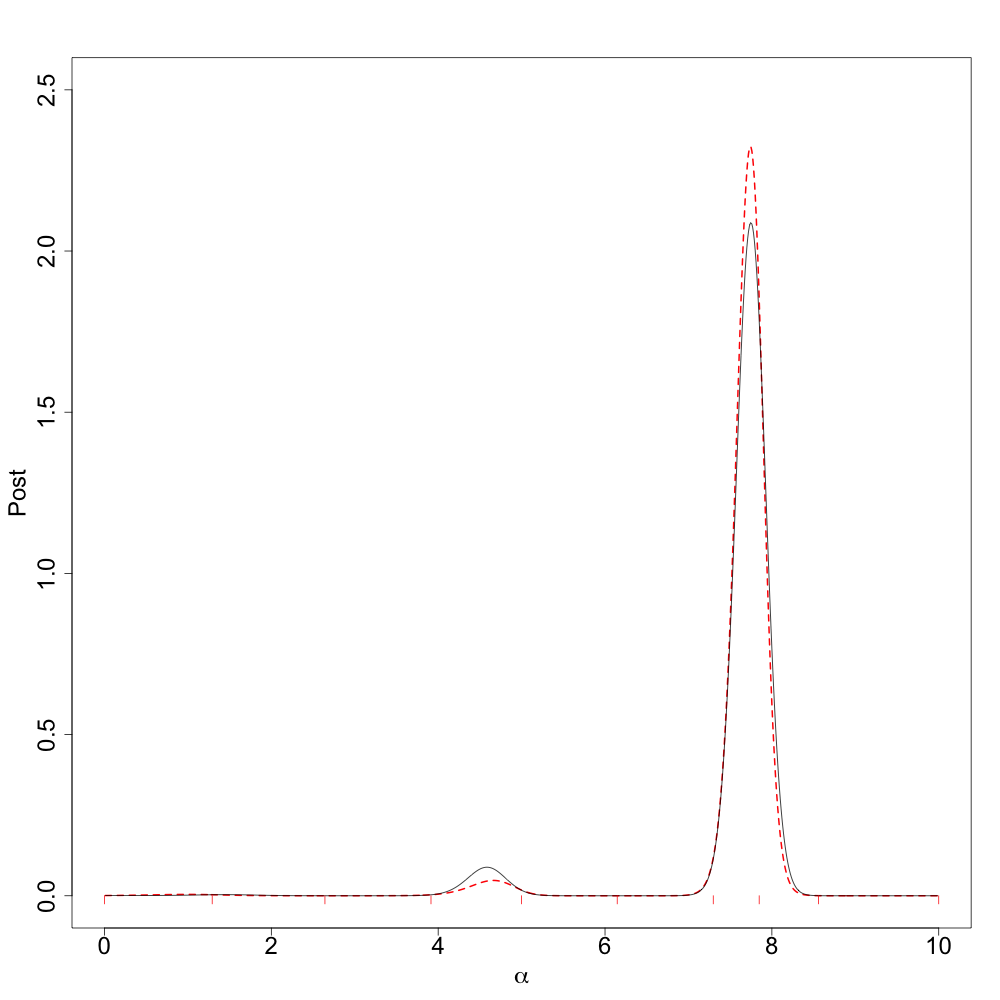}
    }
          \subfigure[Medium Case: $B = 30$]{
      \includegraphics[width=0.31\textwidth]{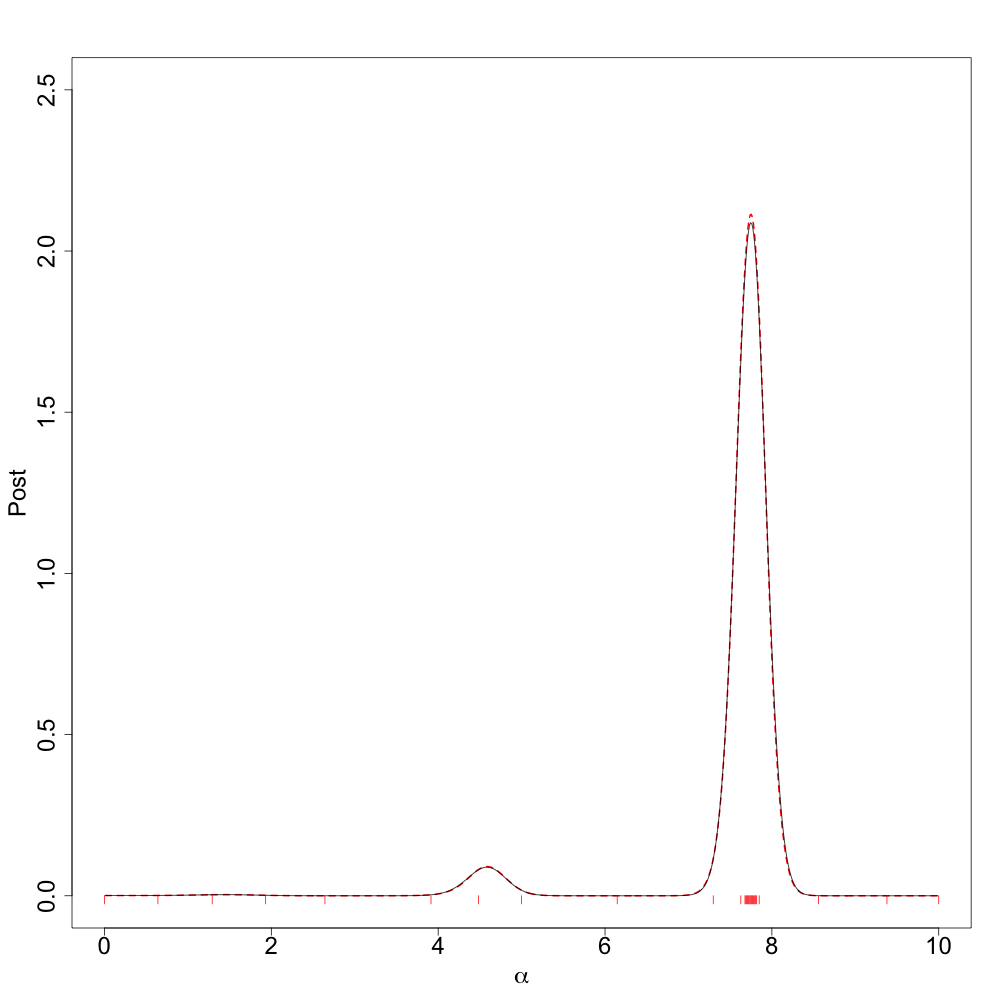}
    }
           \subfigure[Medium Case: $B = 80$]{
      \includegraphics[width=0.31\textwidth]{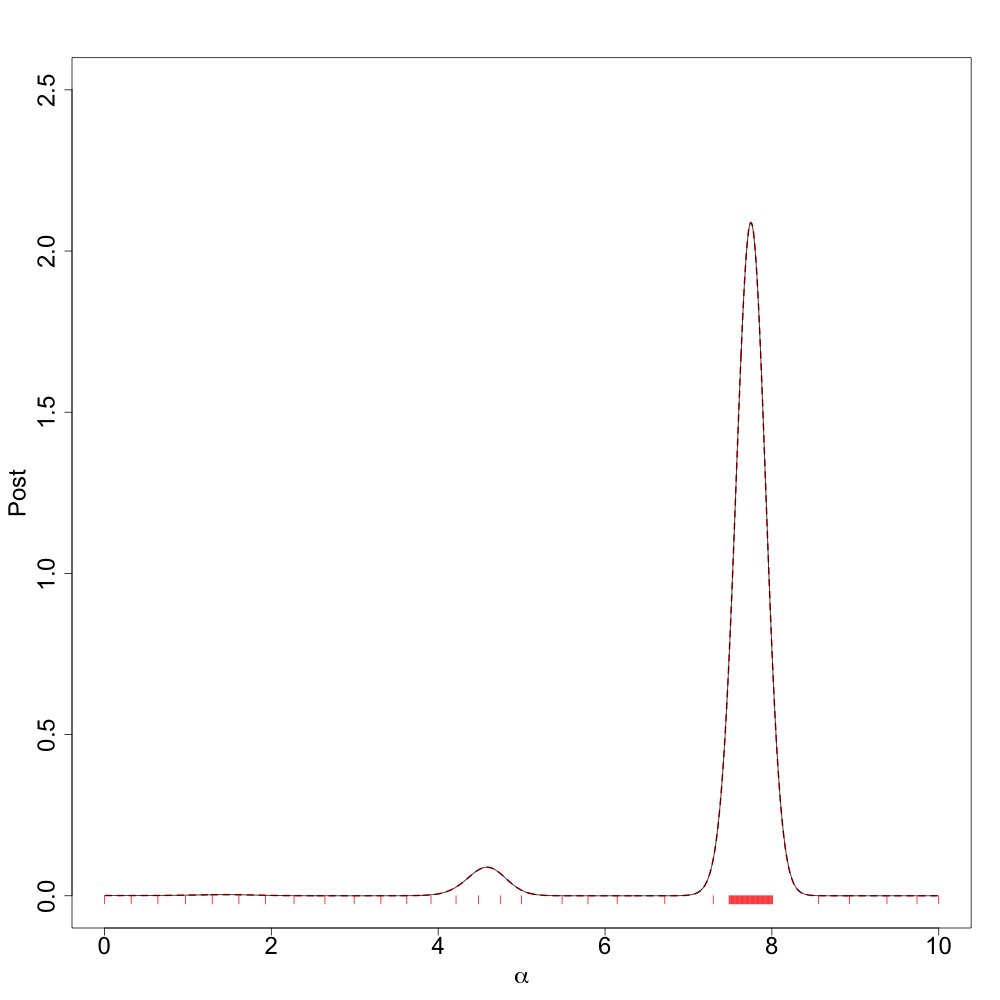}
    }
              \subfigure[Hard Case: $B = 10$]{
      \includegraphics[width=0.31\textwidth]{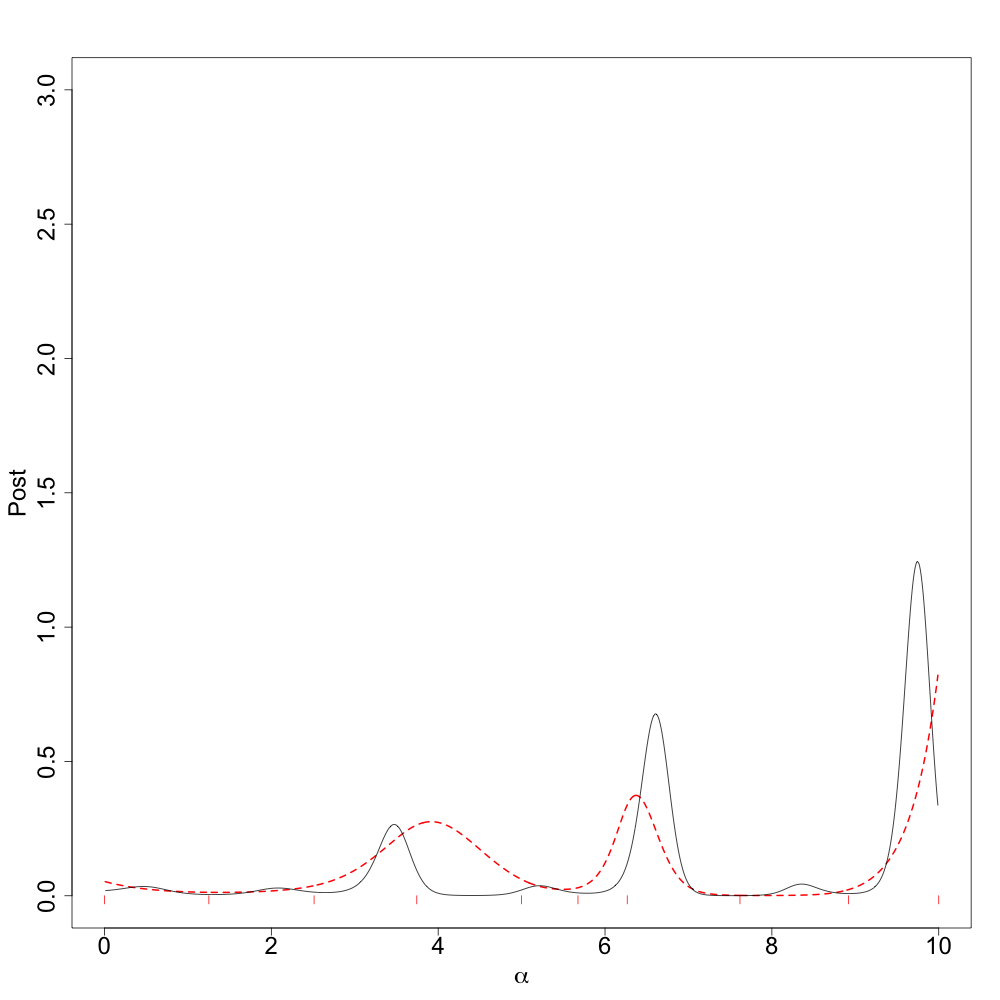}
    }
          \subfigure[Hard Case: $B = 30$]{
      \includegraphics[width=0.31\textwidth]{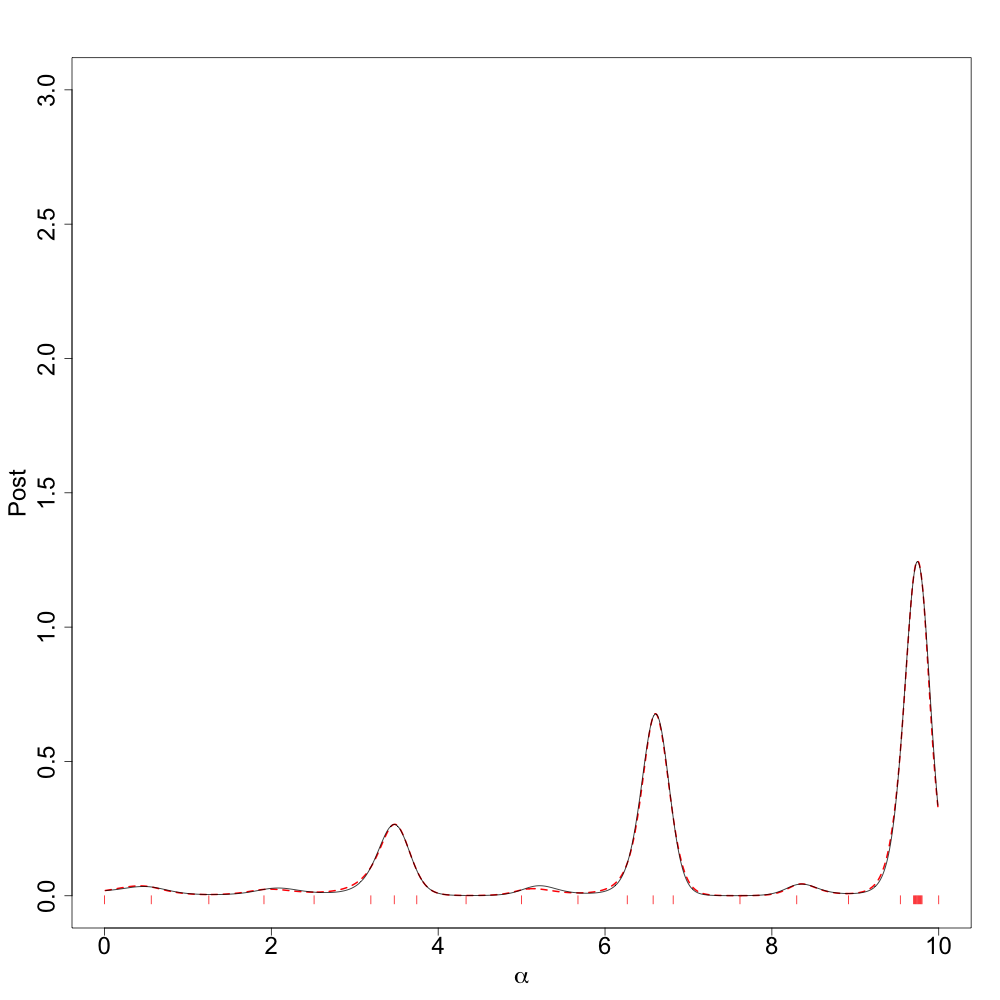}
    }
           \subfigure[Hard Case: $B = 80$]{
      \includegraphics[width=0.31\textwidth]{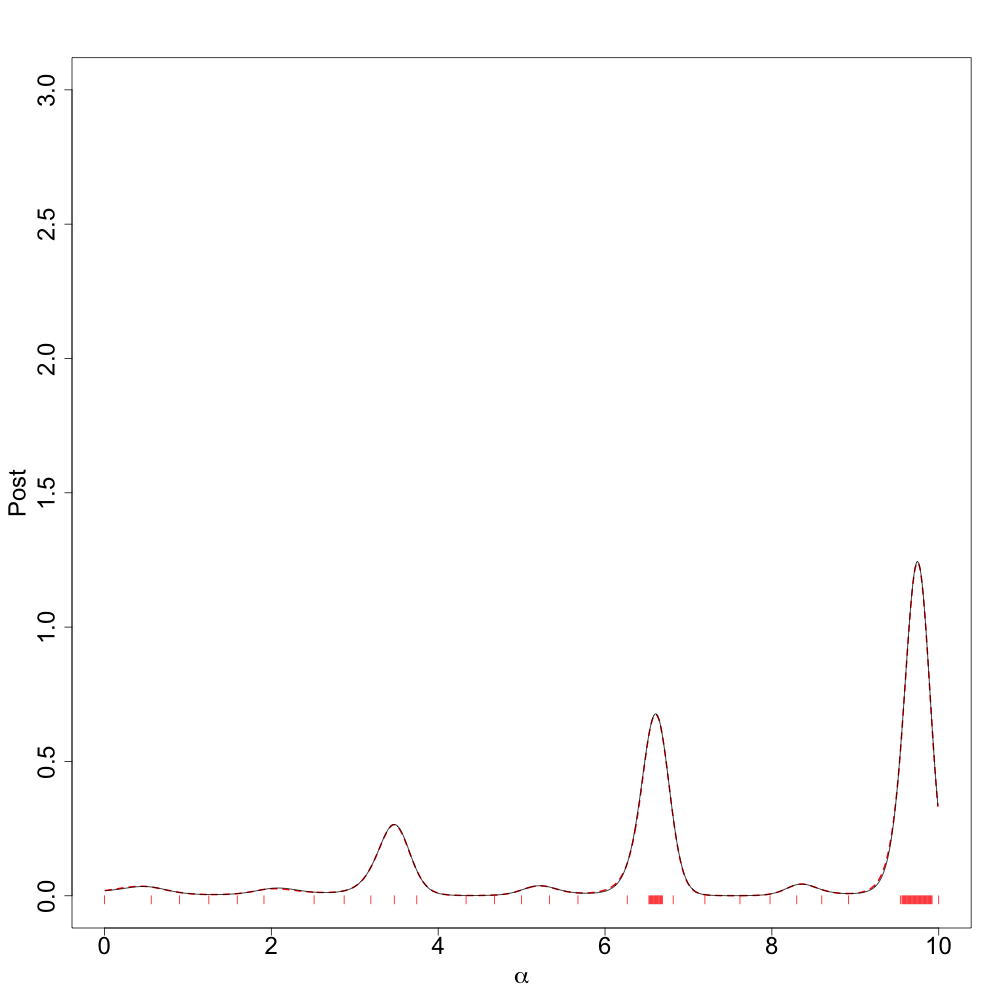}
    }
    
    \caption{Posterior distribution of $\alpha$ in \cref{subsec:simulation1}. The red dashed lines denote the approximated posteriors obtained from the BOSS algorithm using different numbers of BO iterations $B$. The black solid line denotes the true posteriors. The design points of the BOSS algorithm are shown in the red vertical lines on the x-axis.}
    \label{fig:sim1Post}
\end{figure}

The log posteriors ${f}_{\texttt{BO}}(\alpha)$ obtained from the BOSS algorithm with $B = 10, 30, 80$ iterations are compared with the true log posterior $f(\alpha)$, in \cref{fig:sim1logPost}. 
The corresponding (normalized) posteriors $\Tilde{\pi}_{\texttt{BO}}(\alpha|\boldsymbol{y})$ are compared to the truth $\pi(\alpha|\boldsymbol{y})$ in \cref{fig:sim1Post}. 
For the simple and medium case, the BOSS algorithm provides quite accurate approximation with $10$ BO iterations, which are almost identical to the truth. 
For the hard scenario, the BOSS algorithm requires a larger number of BO iterations ($B = 30$) to accurately capture the truth. 
This is expected as the number of local optima in $f(\alpha)$ is larger than the provided number of initial values. In this case, larger number of BO iterations are required for BOSS algorithm to precisely locate the global mode of the function and explore the overall behavior of the function.

\begin{figure}
          \centering
          \subfigure[Simple: KL]{
      \includegraphics[width=0.31\textwidth]{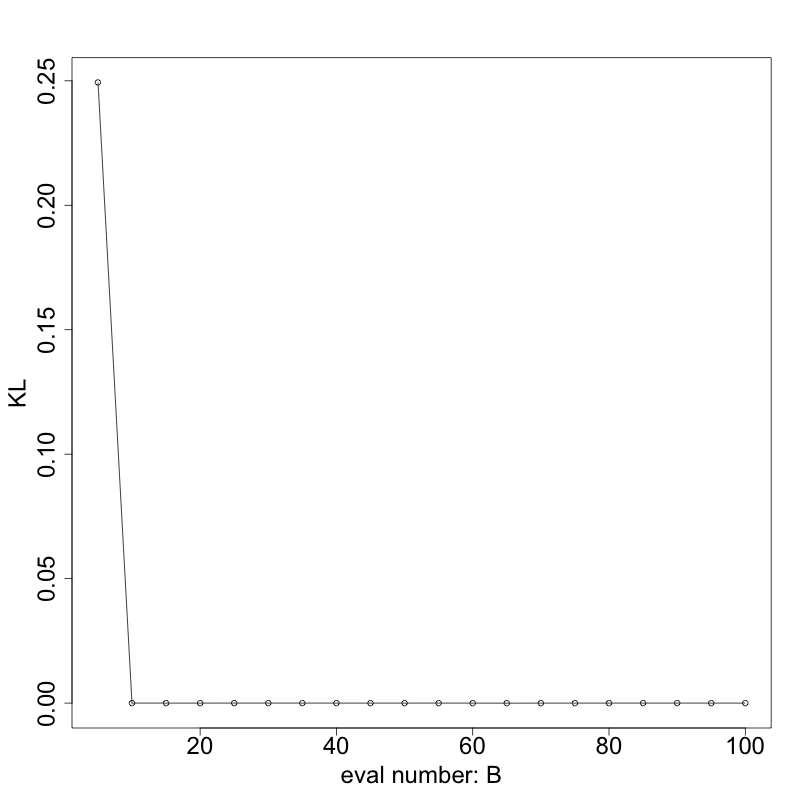}
    }
           \subfigure[Medium: KL]{
      \includegraphics[width=0.31\textwidth]{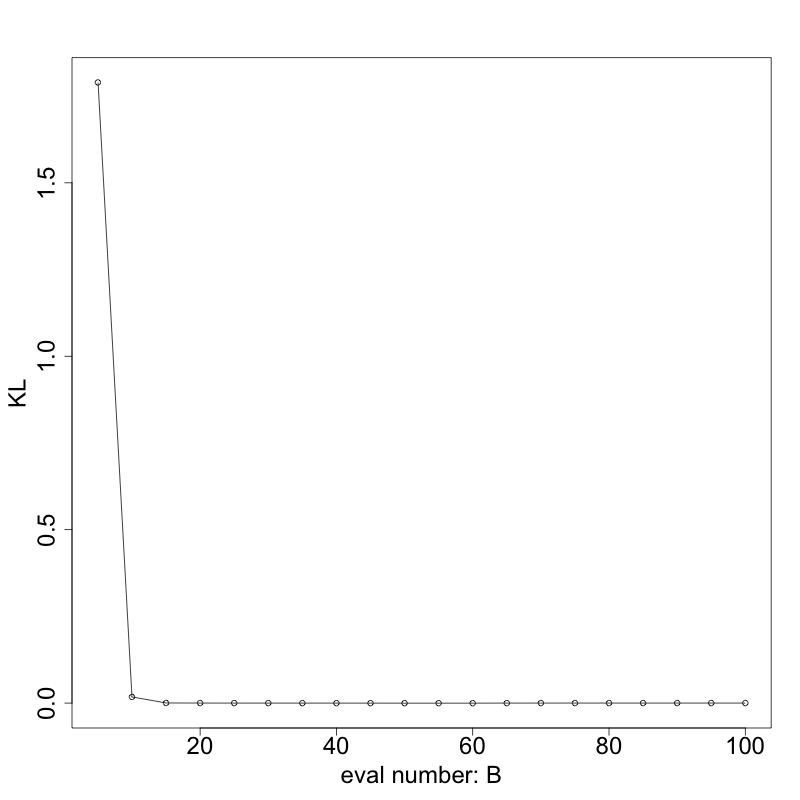}
    }
               \subfigure[Hard: KL]{
      \includegraphics[width=0.31\textwidth]{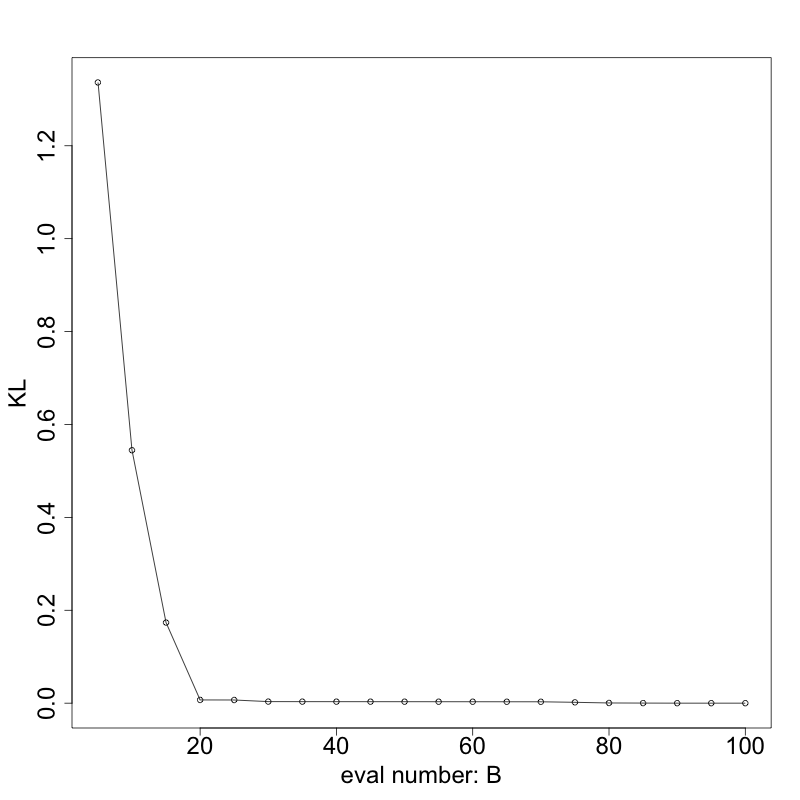}
    }
              \subfigure[Simple: log KS]{
      \includegraphics[width=0.31\textwidth]{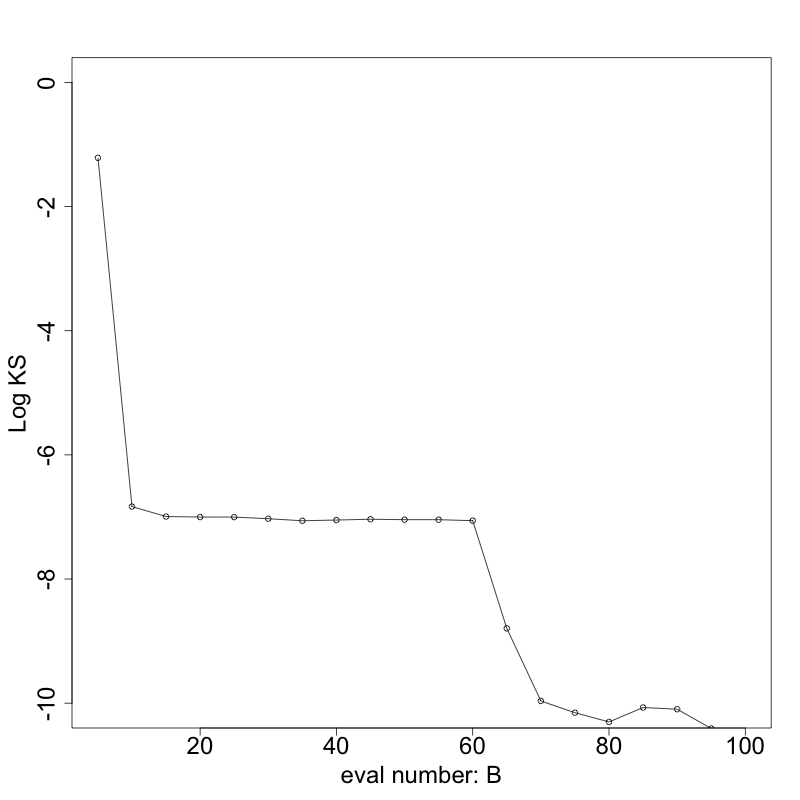}
    }
              \subfigure[Medium: log KS]{
      \includegraphics[width=0.31\textwidth]{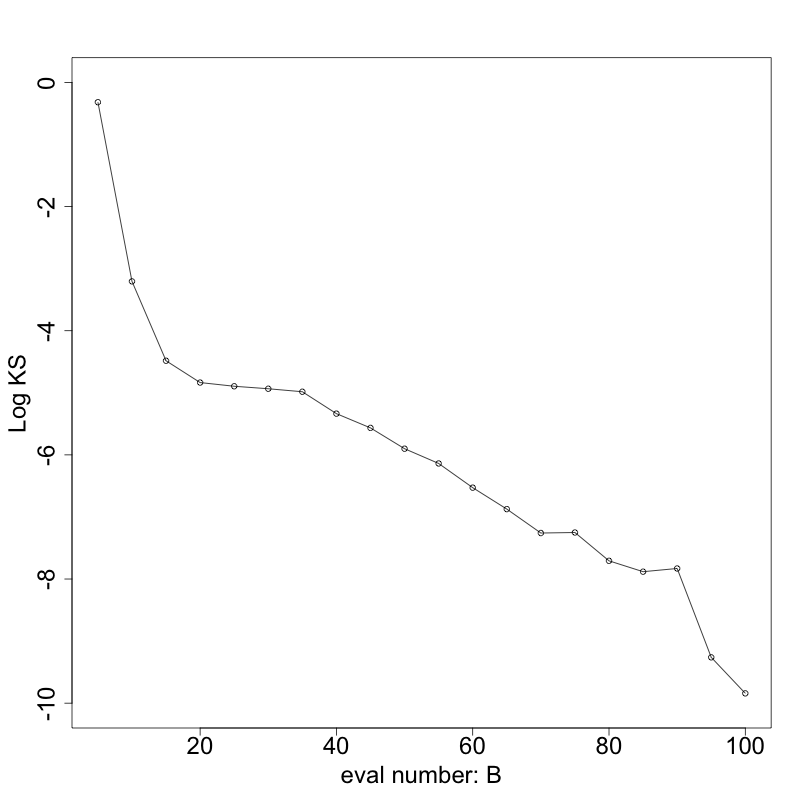}
    }
          \subfigure[Hard: log KS]{
      \includegraphics[width=0.31\textwidth]{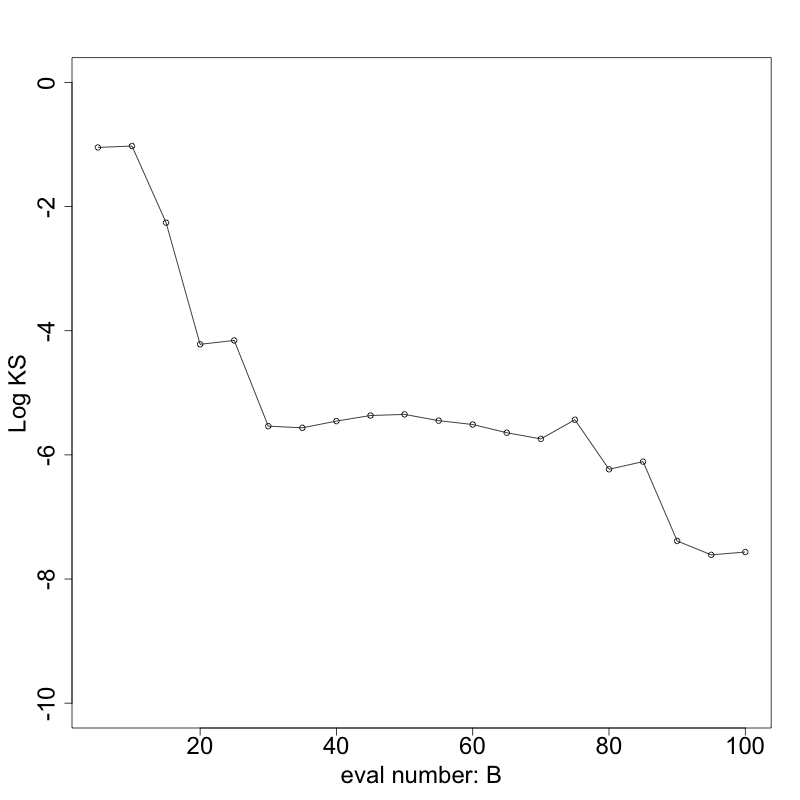}
    }
    \caption{The KL divergence and (log) KS statistic of each BOSS implementation with respect to the number of BO iterations $B$, for the three levels of posterior functions in \cref{subsec:simulation1}.}
    \label{fig:sim1KLKs}
\end{figure}

Finally, to assess the accuracy of BOSS with different numbers of BO iterations, we compute the KL divergence from the approximated posterior $\tilde{\pi}_{\texttt{BO}}(\alpha|\boldsymbol{y})$ to the true posterior $\pi(\alpha|\boldsymbol{y})$, as well as the corresponding Kolmogorov–Smirnov (KS) statistic, which is defined as the largest distance between the two CDFs. 
The KL divergence and (log) KS statistic of each BOSS implementation in each scenario is presented in \cref{fig:sim1KLKs}.
As shown in the figures, both metrics decline fast as the number of iterations increases, illustrating the accuracy of the BOSS approximation.

\subsection{Simulation 2: Inference of an Unknown Periodicity}\label{subsec:simulation2}
In this section, we simulate $n = 100$ observations from the following Poisson regression model:
\[\label{equ:TRUEmodSim2}
\begin{aligned}
    y_i|\mu_i &\sim \text{Poisson}(\mu_i), \\
    \log(\mu_i) =& 1 + 
    0.5 \cos\bigg(\frac{2\pi x_i}{1.5}\bigg) - 1.3  \sin\bigg(\frac{2\pi x_i}{1.5}\bigg) + \\
    & 1.1 \cos\bigg(\frac{4\pi x_i}{1.5}\bigg) + 0.3  \sin\bigg(\frac{4\pi x_i}{1.5}\bigg) + \epsilon_i, \\
    \epsilon_i &\overset{iid}{\sim} \mathcal{N}(0,4).
\end{aligned}
\]
The response variable is denoted as $y_i$, with mean being $\mu_i$.
The covariates $\boldsymbol{x} = \{x_i\}_{i=1}^n$ are equally spaced between $[0,5]$.
The observational level random intercept $\epsilon$ is introduced as the overdispersion.

Given this dataset, we consider the following hierarchical model:
\[\label{equ:ELGMmodSim2}
\begin{aligned}
 y_i|\mu_i &\sim \text{Poisson}(\mu_i), \\
    \log(\mu_i) =& \beta_0 + 
    \beta_1 \cos\bigg(\frac{2\pi x_i}{\alpha}\bigg) - \beta_2  \sin\bigg(\frac{2\pi x_i}{\alpha}\bigg) + \\
    & \beta_3 \cos\bigg(\frac{4\pi x_i}{\alpha}\bigg) + \beta_4  \sin\bigg(\frac{4\pi x_i}{\alpha}\bigg) + \epsilon_i, \\
    \epsilon_i &\overset{iid}{\sim} \mathcal{N}(0,\sigma^2),
\end{aligned}
\]
where independent Gaussian priors $\mathcal{N}(0,1000)$ are assigned to the intercept and all of the four fixed effects. Motivated by \citep{pcprior}, an Exponential prior is used on $\sigma$ such that $P(\sigma > 1) = 0.5$. 

If the periodicity parameter $\alpha$ is known, the model in \cref{equ:ELGMmodSim2} can be fitted as an LGM, with four fixed effects and one random effect. 
However, when $\alpha$ itself is unknown, the standard inferential method no longer works, as the design matrix of the model changes with $\alpha$.
We implement the proposed BOSS algorithm to infer $\alpha$ with five equally spaced initial values and a range of different number of BO iterations $B$. To assess the robustness of BOSS algorithm, we use $\mathcal{N}(3, 0.5^2)$ as the prior for $\alpha$, which is a highly informative prior but mis-specified.
As a comparison, we also use the exact grid method \citep{bivand2014approximate, gomez2018markov}, with a equally spaced grid of size $800$ in the essential support $\Omega = [0.5,4.5]$. 
Given the fine granularity, the result from the grid method is viewed as the oracle in this example.

\begin{figure}[p]
          \centering
          \subfigure[$\pi(\alpha | \boldsymbol{y})$: $B = 15$]{
      \includegraphics[width=0.31\textwidth]{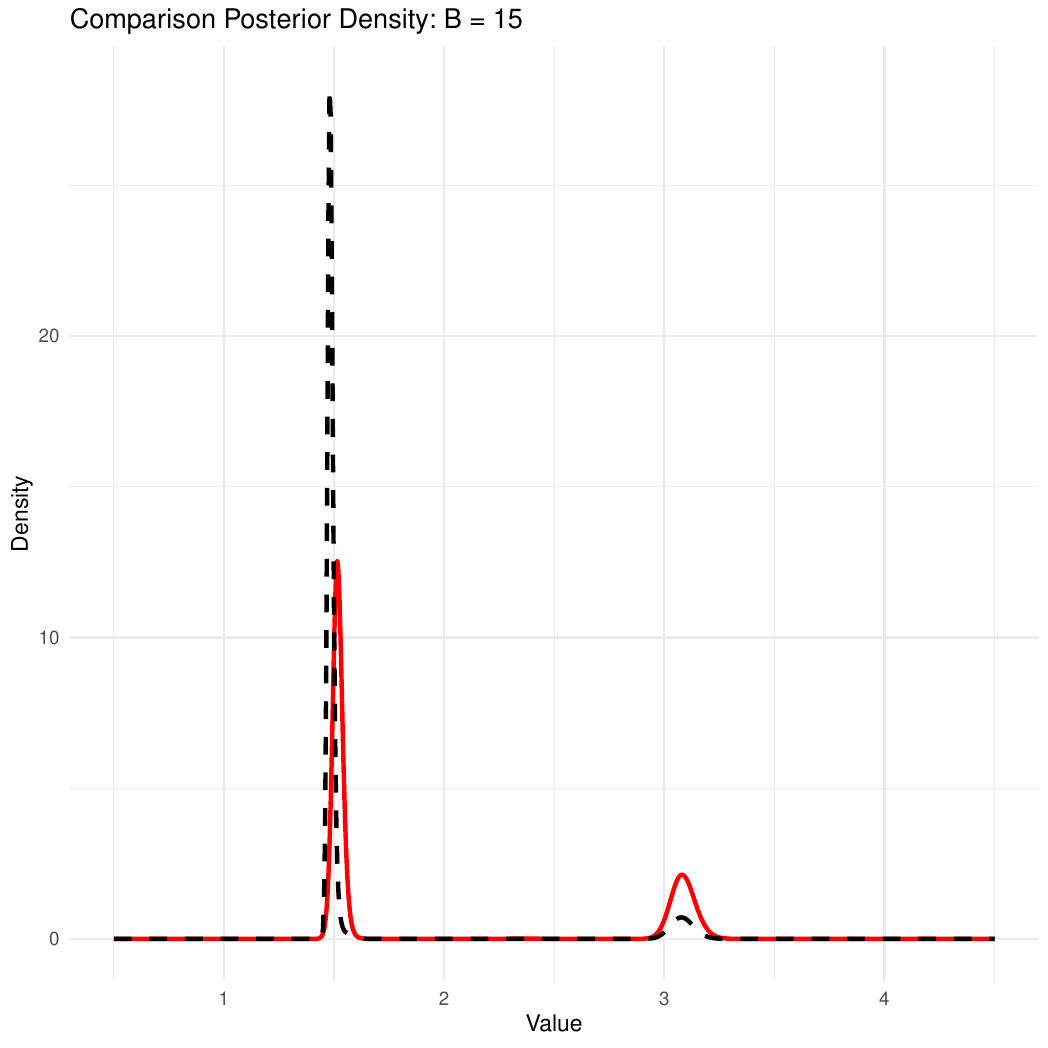}
    }
          \subfigure[$\pi(\alpha | \boldsymbol{y})$: $B = 30$]{
      \includegraphics[width=0.31\textwidth]{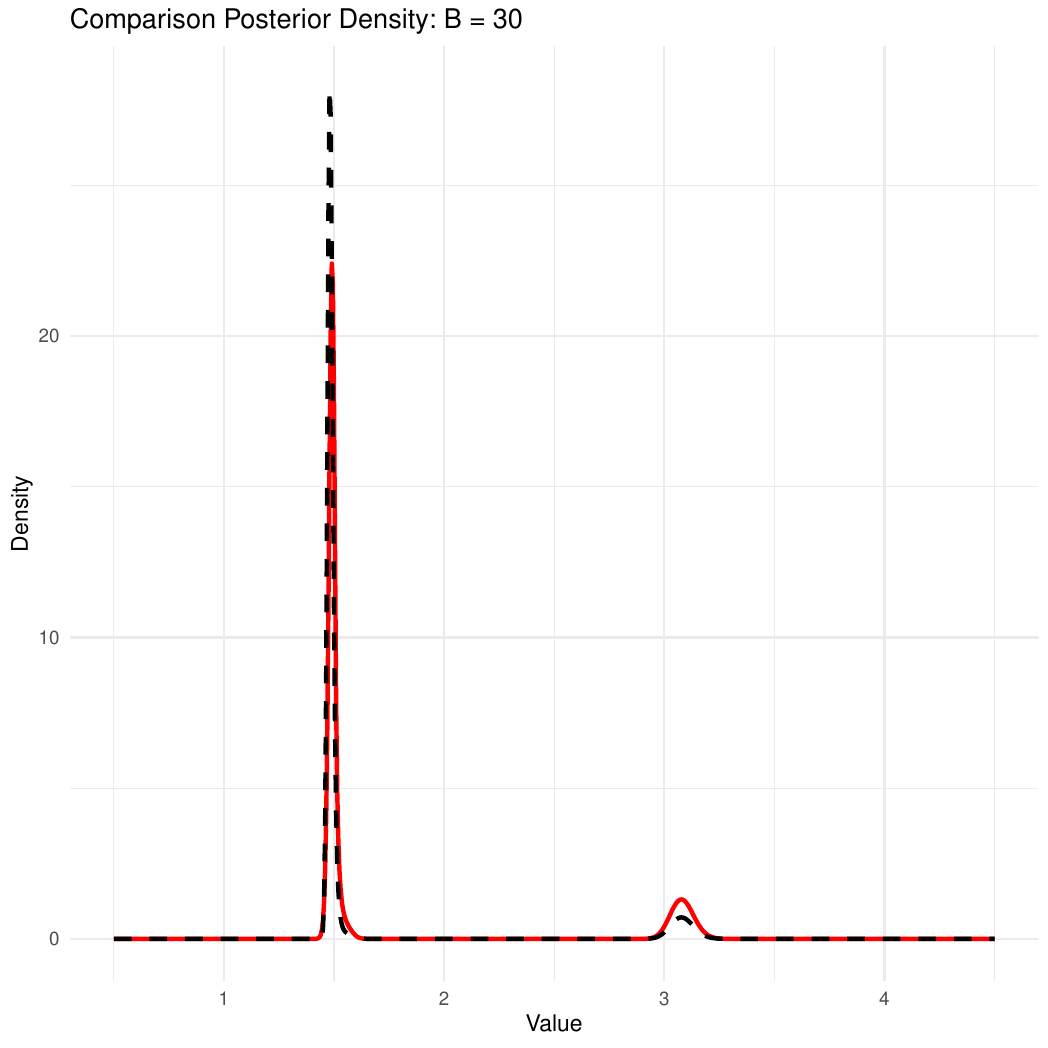}
    }
           \subfigure[$\pi(\alpha | \boldsymbol{y})$: $B = 80$]{
      \includegraphics[width=0.31\textwidth]{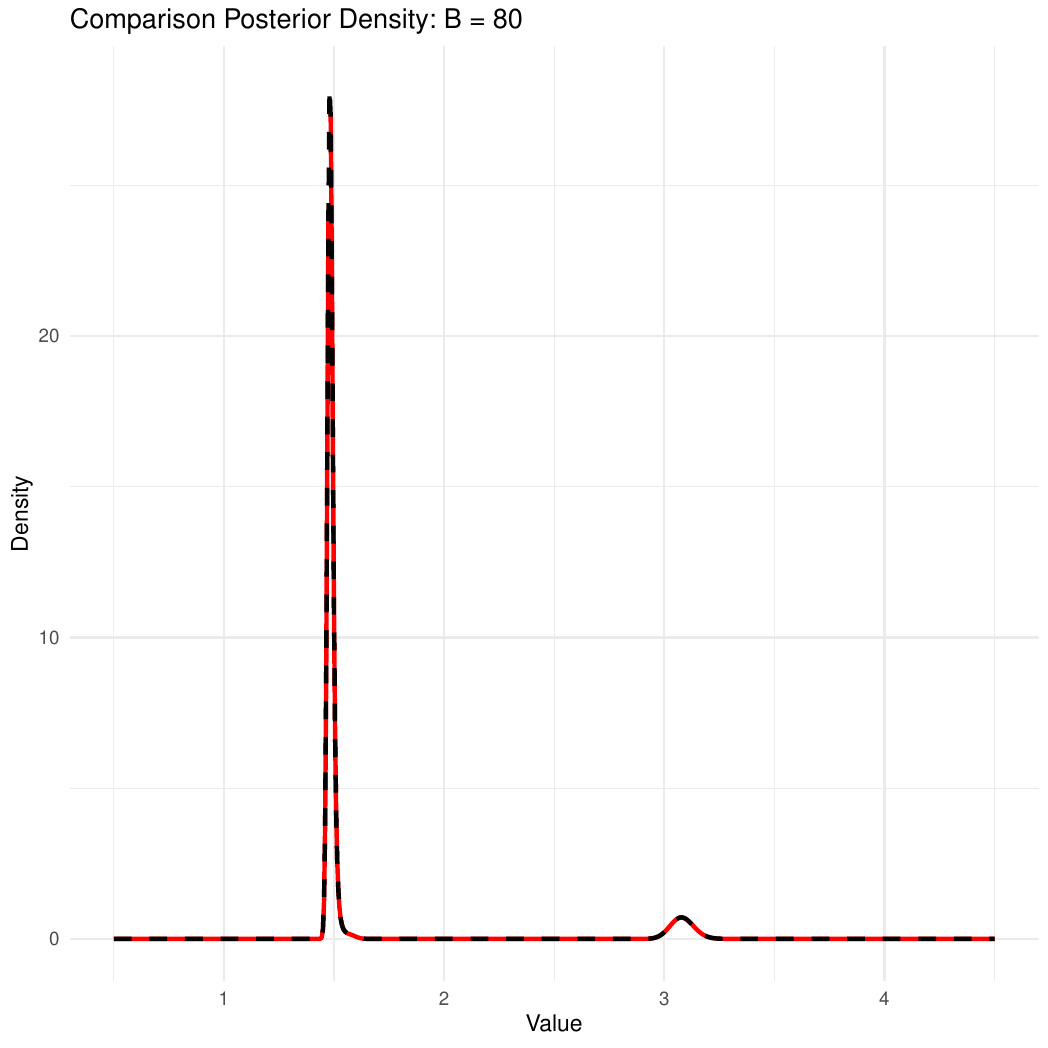}
    }
              \subfigure[Runtime relative to grid method]{
      \includegraphics[width=0.31\textwidth]{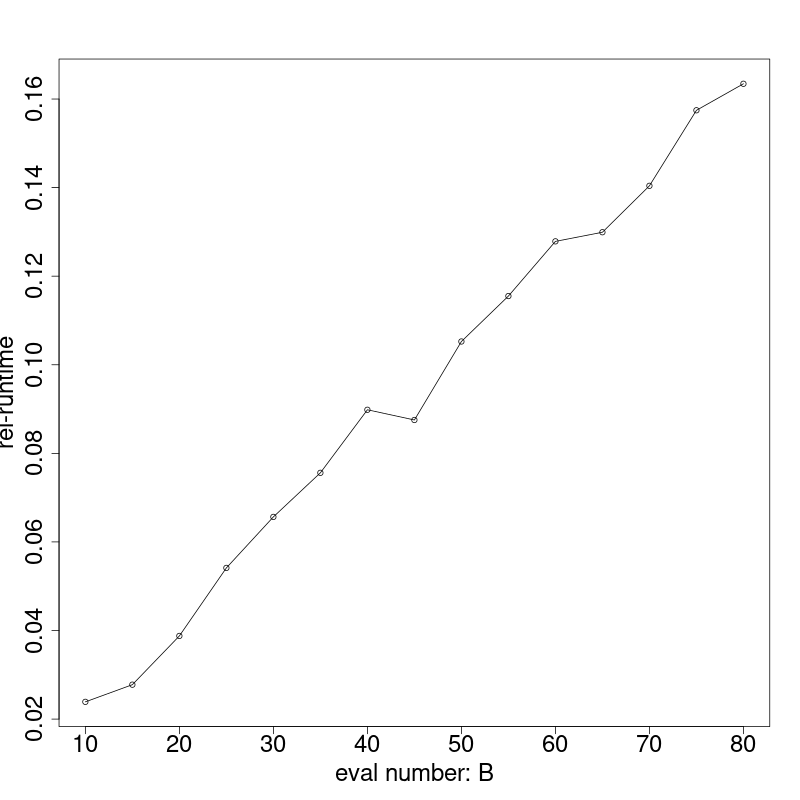}
    }
          \subfigure[KL Divergence]{
      \includegraphics[width=0.31\textwidth]{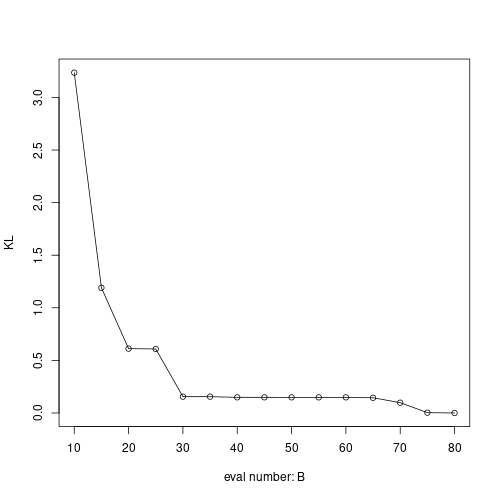}
    }
           \subfigure[KS Distance]{
      \includegraphics[width=0.31\textwidth]{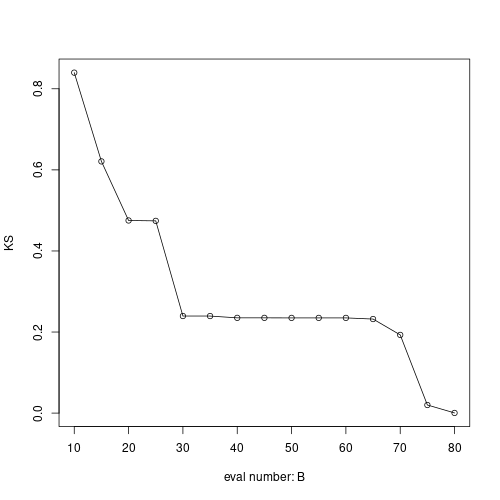}
    }
    
    \caption{(a-c): Posterior distribution of $\alpha$ in \cref{subsec:simulation2}. The red dashed lines denote the approximated posteriors obtained from BOSS with different numbers of BO iterations $B$. The black solid line denotes the oracle posteriors from the grid method.
    (d): The relative-runtime of each BOSS implementation compared to the grid method, as the number of BO iterations $B$ increases. (e-f): The KL divergence and the KS distance of the approximated posterior from BOSS to the posterior from the grid method.}
    \label{fig:sim2Post}
\end{figure}

The results from the proposed BOSS algorithm with different choices of $B$ are shown in \cref{fig:sim2Post}(a-c). 
The approximated posterior from BOSS provides a reasonable estimate of the locations of the two modes with only $B = 15$ iterations. 
When $B = 30$, the approximated posterior correctly captures the majority of the mass of the posterior, with a small underestimation of the density at the global mode. 
Using $B = 80$ iterations, the approximated posterior from BOSS is indistinguishable from the oracle result of the grid method. 
The accuracy of the approximated posteriors can also be confirmed from the plot of its KL divergence and KS distance, comparing to the oracle posterior in \cref{fig:sim2Post}(e-f). 
Finally, we illustrate the significant computational efficiency of BOSS compared to the grid method, by computing its total runtime at each choice of $B$, relative to the total runtime of the grid method. 
As shown in \cref{fig:sim2Post}(d), each additional ten BO iterations only increases the relative runtime by roughly 2 percent. 
With $B = 80$, BOSS produces a posterior that is indistinguishable from the posterior of the oracle method, but only requires a fifth of its total runtime. 
As the granularity of the grid method increases, the computational advantage of BOSS will rapidly grow further.

\subsection{Simulation 3: Change Point Detection}\label{subsec:simulation3}

In this example, we simulate $n = 1000$ observations from the following model with a change point $\alpha = 6.5$:
\[\label{equ:change_point_model}
\begin{aligned}
    y_i|\eta_i &\sim \mathcal{N}(\eta_i, \sigma^2), \\
    \eta_i &= g_1(x_i)\mathbb{I}(x_i \leq \alpha) + g_2(x_i)\mathbb{I}(x_i > \alpha), \\
    g_1(x) &= x \log(x^2 + 1), \quad g_2(x) = 3.3 x + 3.035.
\end{aligned}
\]
The response variable is denoted by $y_i$ with mean $\eta_i$ and standard deviation $\sigma = 0.3$.
The covariate $x_i$ is equally spaced in the interval $[0,10]$. 
We assume $g_1$ and $g_2$ are two unknown smooth functions, continuously joined at an unknown change point $\alpha$. 
The target of inference is the posterior of $\alpha$, as well as the two unknown functions.

For priors, we assign two independent second order Integrated Wiener process ($\text{IWP}_2$) priors to $g_1$ and $g_2$, respectively with standard deviation (SD) parameters $\sigma_1$ and $\sigma_2$. 
A uniform prior is assigned to the change point $\alpha$ with support $\Omega = [0,10]$. 
For the SD parameters $\sigma_1$, $\sigma_2$ and $\sigma$, we use independent Exponential priors with median $1$. 
To facilitate the computation, we use the O-Spline approximation in \cite{IwpOsplines} with $100$ equally spaced basis functions for each IWP prior.
Given a fixed value of $\alpha$, the model in \cref{equ:change_point_model} can be readily fitted as an ELGM or a LGM through inference methods described in \cite{elgm} and \cite{inla}. However, when $\alpha$ is unknown, the inference becomes much more involved, as the partition of the effect into the two functions $g_1$ and $g_2$ becomes unknown. 
Therefore, we employ the BOSS algorithm proposed in the earlier section, with $\alpha$ being the conditioning variable. 
As comparisons, we also implement the MCMC method \citep{gomez2018markov}, and the exact grid method \citep{bivand2014approximate, gomez2018markov}. 
The MCMC method yields $3000$ posterior samples after burn-in ($1000$ samples) and thinning (1 per 3 samples). 
The grid method uses a fine grid of $1000$ equally spaced values in the support of the prior, and then is interpolated with a smoothing spline. 
Given the fine granularity, the result from the grid method is treated as the oracle in this comparison.

\begin{figure}
    \centering
           \subfigure[BO with 20 iterations:]{
      \includegraphics[width=0.45\textwidth]{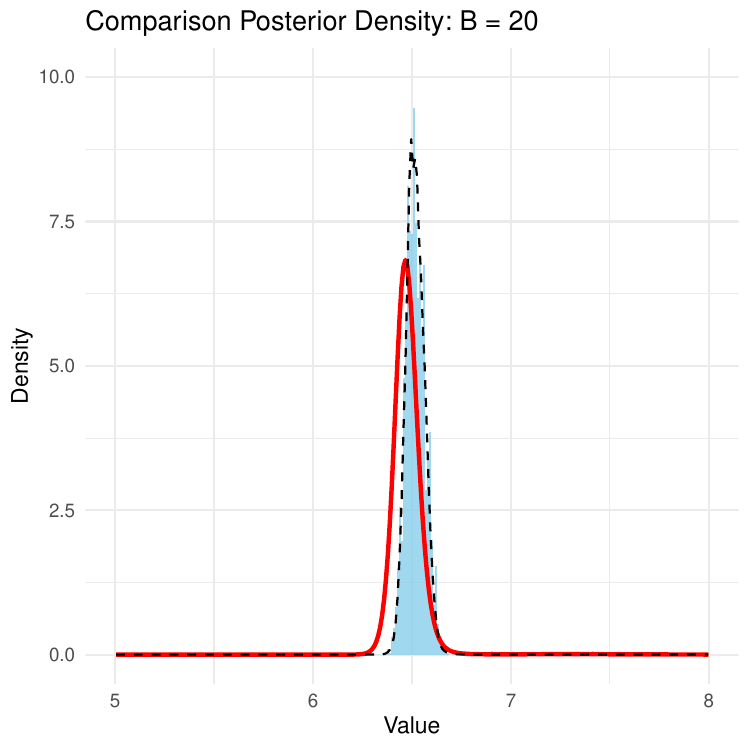}
    }
            \subfigure[BO with 30 iterations:]{
      \includegraphics[width=0.45\textwidth]{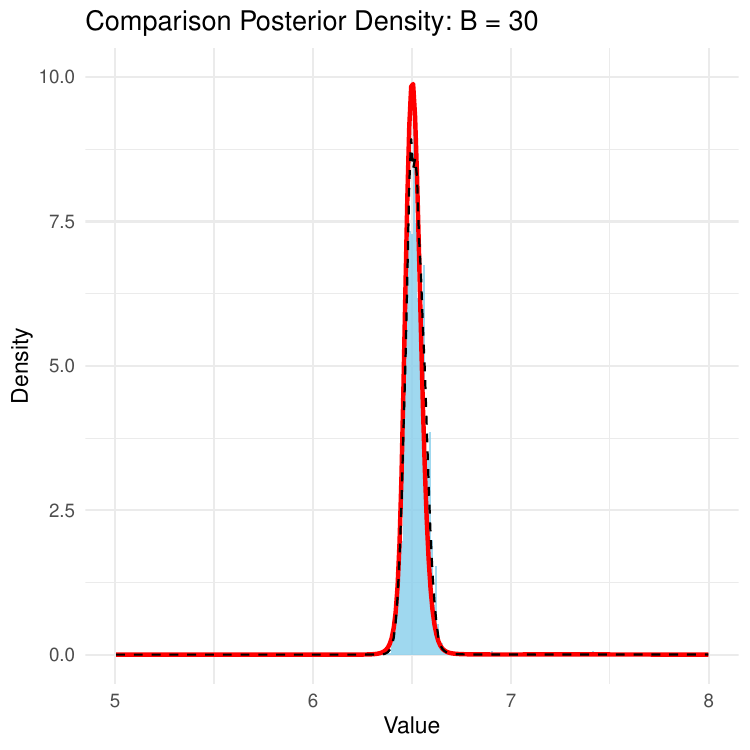}
    }
    \caption{Posterior distribution of the change point $\alpha$ in \cref{sec:simulation}. The red solid lines denote the approximated posteriors obtained from the BOSS algorithm using different numbers of BO iterations. The blue histogram denotes the posterior samples obtained from the MCMC method. The dashed line denotes the posterior obtained using the grid method.}
    \label{fig:change-point-BO}
\end{figure}

As shown in the \cref{fig:change-point-BO}, the proposed BOSS algorithm produces approximation to the posterior distribution with comparable accuracy to the MCMC and the grid methods. 
When the number of BO iteration is 20, the BO approximation has a slight bias in the location, but this bias can be effectively corrected by running another 10 iterations of BO. 
In terms of the computational runtime, the BOSS algorithm only evaluates the true joint density $\pi(\alpha, \boldsymbol{y})$ for $B = 20$ or $30$ times, whereas the fine-grid method and the MCMC method respectively evaluate this density for $1000$ and $10,000$. 
As a result, the BOSS algorithm is able to produce comparable posteriors to MCMC and fine-grid method, using just a fraction of their runtimes.

\begin{figure}
    \centering
            \subfigure[$g_1$]{
      \includegraphics[width=0.41\textwidth]{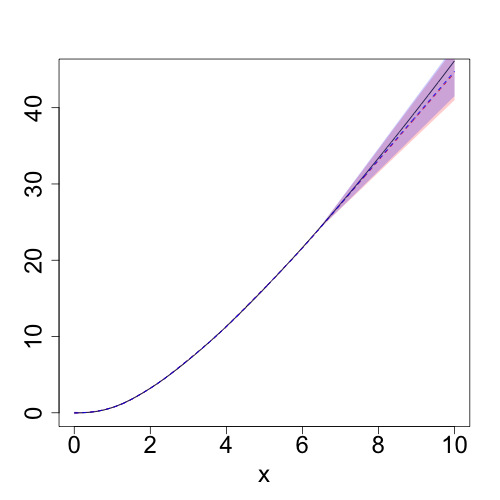}
    }
           \subfigure[$g_2$]{
      \includegraphics[width=0.41\textwidth]{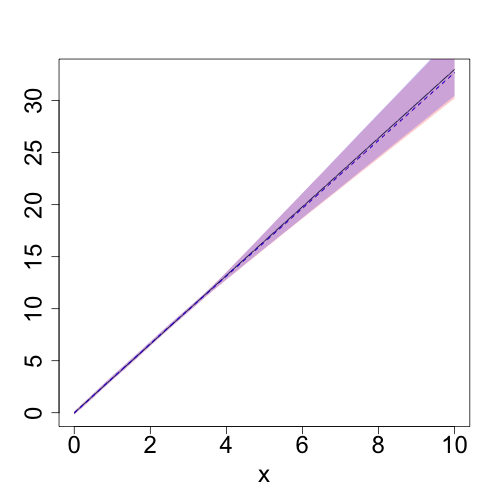}
    }
    \caption{Posterior distribution of the two unknown functions in \cref{sec:simulation} obtained from the BOSS (blue) and the oracle grid method (red). The 95 percent credible intervals are shown as the shaded regions; the posterior medians are shown as the dashed lines; and the true functions are shown as the black solid lines.}
    \label{fig:change-point-BO-function}
\end{figure}

To obtain the posterior inference of the two unknown functions $g_1$ and $g_2$, we compute the AGHQ rule on the BOSS surrogate $\tilde{\pi}_{\texttt{BO}}(\alpha|\boldsymbol{y})$ with $B = 30$ iterations, using $K = 10$ quadrature points.
The details of this procedure are provided in \cref{subsec:BOSS_latent}.
Given $\tilde{\pi}_{\texttt{BO}}(\alpha|\boldsymbol{y})$ is an analytical function with no evaluation cost, the AGHQ rule $\mathcal{Q}(K)$ can be computed efficiently on this surrogate function. The only additional computational cost is due to fitting the additional $K = 10$ LGMs with $\alpha$ fixed at each of the quadrature points. 
As a comparison, we also apply the same AGHQ method on the oracle posterior of $\alpha$ obtained from the fine-grid interpolation, and obtain the posterior of the two functions with this oracle approach. 
As shown in \cref{fig:change-point-BO-function}, the posteriors obtained from the BOSS method are indistinguishable from the posterior obtained from the oracle fine-grid approach.

\subsection{Simulation 4: Non-Linear Regression}

In this example, we consider the following non-linear regression with two nuisance parameters $R > 0$ and $\beta > 0$: 
\[\label{eqn: log plummer}
y_i \mid \log\rho(r_i) &\sim \mathcal{N}(\log\rho(r_i), \sigma^2), \\
\log\rho(r_i) &= \log\rho_0 - \gamma\log\left\{1 + (r_i/R)^\beta\right\}.
\]
The response variable is denoted as $y_i$ with mean $\log \rho(r_i)$ and SD $\sigma$, and the covariate is denoted as $r_i$.  
The \cref{eqn: log plummer} is widely used in astrophysics \citep{plummer_problem_1911, evans_hypervirial_2005, de_rijcke_dynamics_2014, jeffreson_gaiaeso_2017} to model the observed radial density ($y_i$) of various celestial objects, such as the density of stars in star clusters: at the center of a star cluster ($r_i = 0$), the theoretical density $\rho(0) = \rho_0$ of stars is the highest; as one moves away from the center of the star cluster, the theoretical density drops according to $\rho(r_i)$ as a function of the distance $r_i$ to the center of the star cluster. A classical example is when $\beta = 2$ and $\gamma = 2.5$, which is the Plummer model \citep{plummer_problem_1911}. Accurately inferring the parameters from data is crucial to help astrophysicists test the validity of their theories on how star clusters form.

The model \ref{eqn: log plummer} reduces to a simple linear regression if the two nuisance parameters $\bm{\alpha} = (R, \beta)$ are known. When the two nuisance parameters are unknown, this model could not be fitted as an LGM using the existing approximate Bayesian inference method as the design matrix varies with the nuisance parameters. In astrophysics, a grid-based approach is usually adopted to first fix $\bm{\alpha}$ at its optimal value, and then make subsequent inference on $\rho_0$ and $\gamma$. This approach ignores the uncertainty associated to the two nuisance parameters, hence could not yield reliable uncertainty quantification. An alternative approach is to make fully Bayesian inference based on the MCMC method, which accounts for the uncertainty quantification of the nuisance parameters at the cost of
longer computational time. More recently, a new method was introduced to facilitate such Bayesian non-linear regression via INLA implemented in the \texttt{R} package \texttt{inlabru} \citep{Yuan_2017, Bachl_2019}. \texttt{inlabru} carries out a first-order Taylor expansion of the non-linear function with respect to the nuisance parameters so that the model can be fitted using INLA as a LGM.

To illustrate the practical utility and inferential accuracy of the proposed BOSS method, we simulate mock data with $n = 201$ according to \ref{eqn: log plummer} with parameters $\rho_0 = 10$, $R = 2$, $\beta = 2$, $\gamma = -2.5$, and $\sigma = 0.5$. 
For these parameters, we assume the following independent priors
\[
\rho_0 \sim \mathcal{N}(0, 1000), & \ R \sim \text{Unif}(0.1, 5), \\
\beta \sim \text{Unif}(0.1, 4), \ \gamma  \sim \mathcal{N}(0, & \ 1000), \ \sigma^2 \sim \text{Inv-Gamma}(1, 10^{-5}).
\]
The conditioning parameters of the BOSS are $\boldsymbol{\alpha} = (R, \beta)$, and the BOSS algorithm is run with 100 BO-iteration, with the log marginal likelihood $\log\pi(\boldsymbol{y} \mid \boldsymbol{\alpha})$ at each iteration obtained from INLA \citep{inla}.
As comparisons, we also make inference using the MCMC and \texttt{inlabru}. 
For the MCMC, four independent chains are run using the No-U-Turn Sampler (NUTS) in \texttt{stan} to obtain $49500$ sample after burn-in ($1000$) and thinning (1 per 8 samples). 
Because of the large number of iterations, the result from this MCMC method is treated as the oracle in this example.

\begin{figure}[H]
\centering
          \subfigure[Joint Posterior of $(R, \beta)$ from BOSS algorithm]{
      \includegraphics[width=0.45\textwidth]{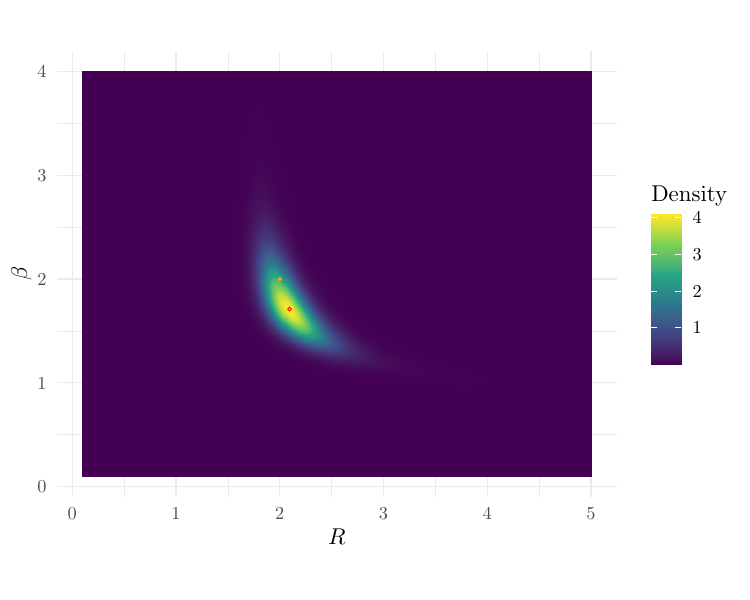}
    }
          \subfigure[Joint Posterior of $(R, \beta)$ from MCMC]{
      \includegraphics[width=0.45\textwidth]{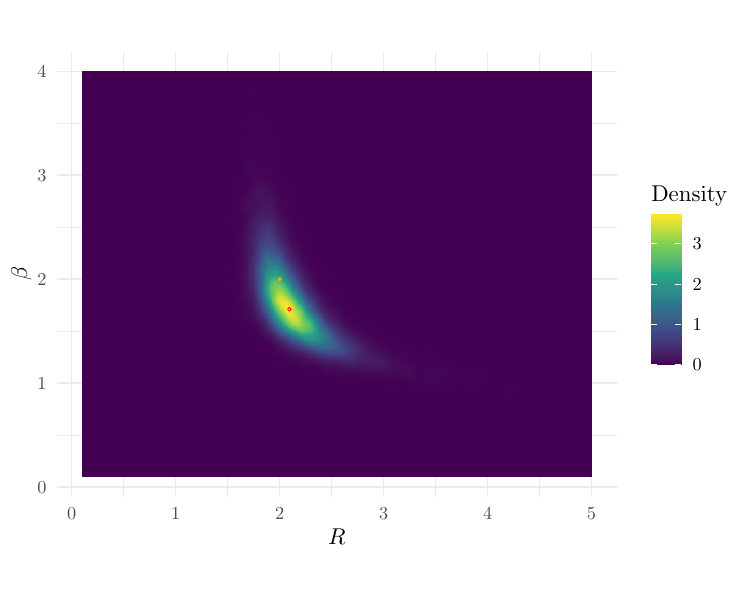}
    }
    \subfigure[Joint Posterior of $(R, \beta)$ from \texttt{inlabru}]{
      \includegraphics[width=0.45\textwidth]{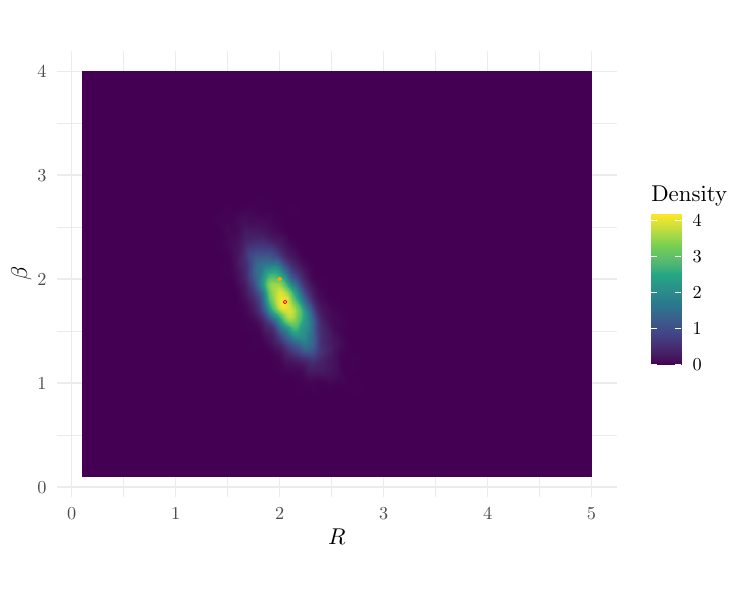}
    }
    \caption{Joint posterior distributions obtained from (a) our BOSS algorithm, (b) MCMC, and (c) \texttt{inlabru}. Orange point is the true parameter value of $(R, \beta) = (2, 2)$. Red circles are the posterior mode obtained from each of the three respective algorithms.}
    \label{fig:joint R beta}
\end{figure}

\begin{figure}[H]
\centering
          \subfigure[Marginal Posterior of $R$]{
      \includegraphics[width=0.45\textwidth]{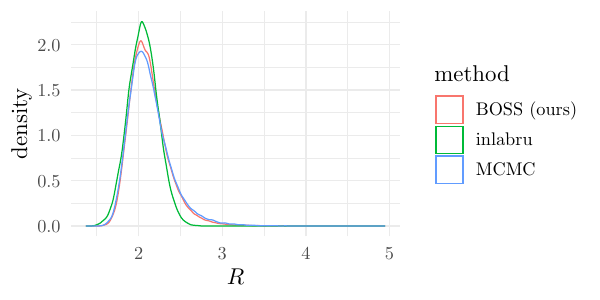}
    }
          \subfigure[Marginal Posterior of $\beta$]{
      \includegraphics[width=0.45\textwidth]{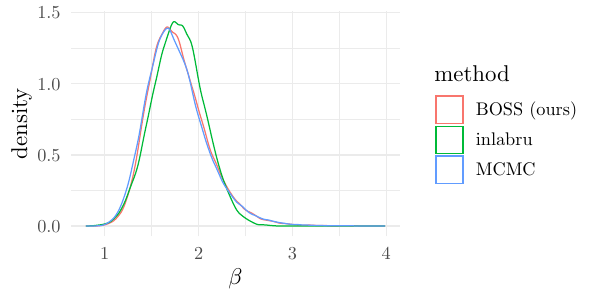}
    }
    \caption{Posterior marginal distributions obtained from our BOSS algorithm (red), \texttt{inlabru} (green), and MCMC (blue) for $R$ and $\beta$.}
    \label{fig:marginal R beta}
\end{figure}

\cref{fig:joint R beta} shows the joint posterior distribution of $(R, \beta)$ obtained form our BOSS algorithm, MCMC, and \texttt{inlabru}.
From \cref{fig:joint R beta}, we see that the joint posterior obtained from BOSS is nearly identical to the posterior obtained from MCMC, demonstrating the inferential accuracy of the proposed approach. However, \texttt{inlabru} completely mis-characterizes the joint posterior distribution due to the linear approximation applied towards the non-linear predictor, and it only produces a Gaussian approximation of the true joint posterior.
Figure \ref{fig:marginal R beta} demonstrates the posterior marginal distributions of $R$ and $\beta$ from our BOSS algorithm, MCMC, and \texttt{inlabru}. For a fair comparison, $49500$ samples are generated based on the results of BOSS and \texttt{inlabru}.
Based on \cref{fig:marginal R beta}, we see that the marginal posterior distributions for both $R$ and $\beta$ obtained from BOSS are comparable to those from MCMC. 
On the contrary, both marginal posterior distributions from \texttt{inlabru} show significant deviation from the ones obtained from MCMC and from BOSS. 
This is because \texttt{inlabru} constructs a linear approximation of the non-linear predictor at a fixed location, and then considers a Laplace approximation at the posterior mode of the nuisance parameters. 
Thus, if the function is highly non-linear, \texttt{inlabru} may only be reliable for estimating the posterior mode but not for accurately characterizing the entire posterior distribution. 
On the contrary, the BOSS algorithm resolves this issue through an efficient exploration of the posterior guided by BO.

\section{Applications}\label{sec:applications}

\subsection{Change Points in All-Cause Mortality}\label{subsec:mortality}
In this example, we consider the analysis of the weekly all-cause mortality counts in the Netherlands and Bulgaria. 
The all-cause mortality data is obtained from the \textit{World Mortality Dataset} \citep{karlinsky2021tracking}, which contains the country-level weekly death counts from 2015 to 2022.

The analysis of all-cause mortality could reveal how much extra mortality was caused by a type of disease or event in a certain time period, relative to the expected mortality using the pre-event data. For example, \cite{knutson2023estimating} and \cite{msemburi2023estimates} have studied how much excess mortality were caused by COVID-19 from 2020 to 2022, with expected mortality determined using the data before 2020. However, the assumption that all the countries had COVID-19 spread-out at January 1st, 2020 is likely unrealistic. 
Hence it is important to have accurate and efficient inference of the change point of the all-cause mortality in each country. 

To make inference of the change-point, we consider the following model for the weekly-mortality:
\begin{equation}
\begin{aligned}
    y(t_i)|\mu(t_i) &\sim \text{Poisson}(\mu(t_i)), \\
    \log(\mu(t_i)) &= 
\begin{cases} 
    g_{tr,\text{pre}}(t_i) + g_{s,\text{pre}}(t_i) & \text{if } t_i \leq a, \\
    g_{tr,\text{pos}}(t_i) + g_{s,\text{pos}}(t_i) & \text{if } t_i > a.
\end{cases}\\
g_{tr,\text{pre}}(t_i) \sim &\text{IWP}_2(\sigma_{tr,\text{pre}}), \ g_{tr,\text{pos}}(t_i) \sim \text{IWP}_2(\sigma_{tr,\text{pos}}), \\
g_{s,\text{pre}}(t_i) \sim &\text{sGP}(\sigma_{s,\text{pre}}), \ g_{s,\text{pos}}(t_i) \sim \text{sGP}(\sigma_{s,\text{pos}}).
\end{aligned}
\end{equation}

\begin{figure}
    \centering
    \subfigure[$\tilde{\pi}_{\texttt{BO}}(\alpha|\boldsymbol{Y})$ in NL]{
        \includegraphics[width=0.4\textwidth]{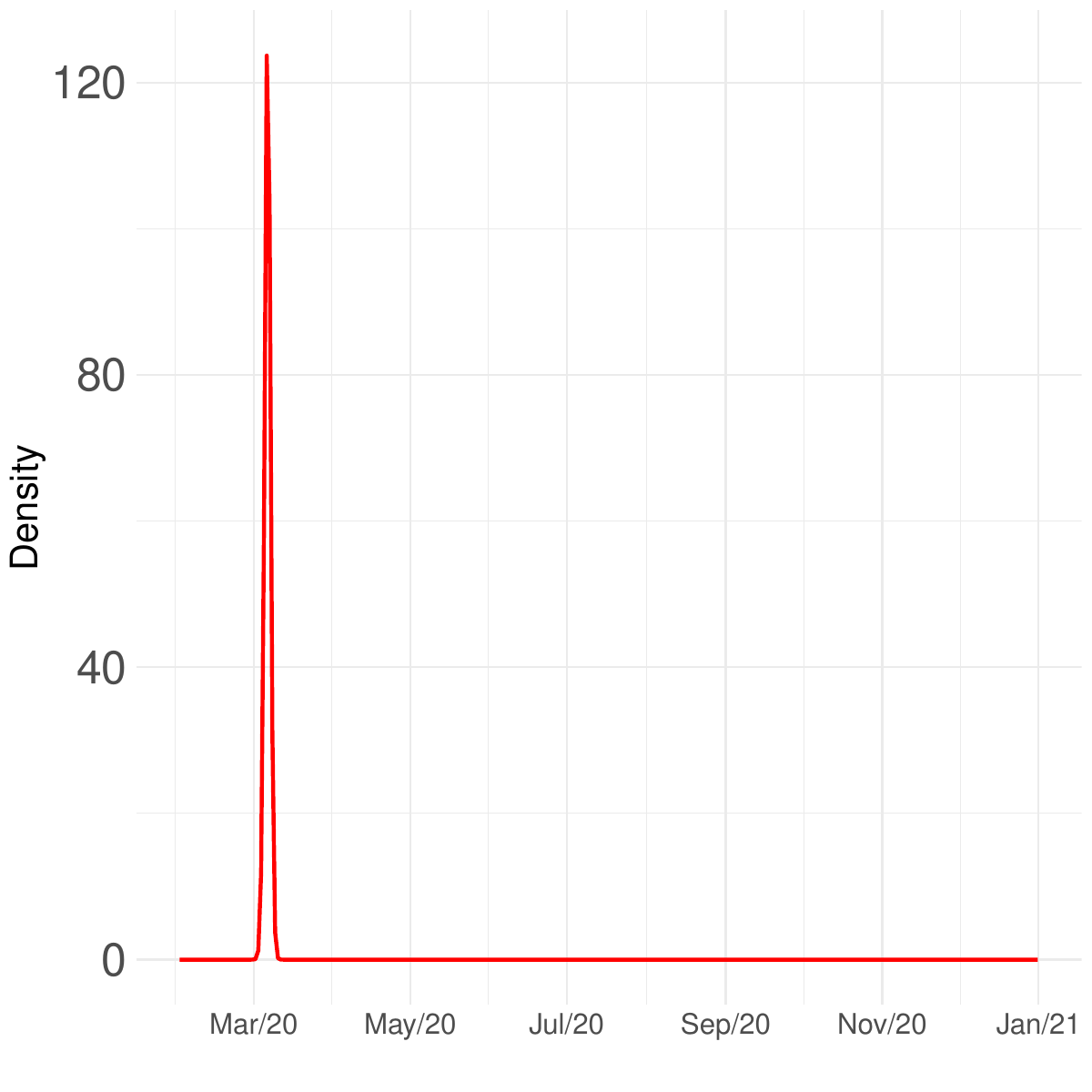}
    }
        \subfigure[$\tilde{\pi}_{\texttt{BO}}(\alpha|\boldsymbol{Y})$ in BG]{
        \includegraphics[width=0.4\textwidth]{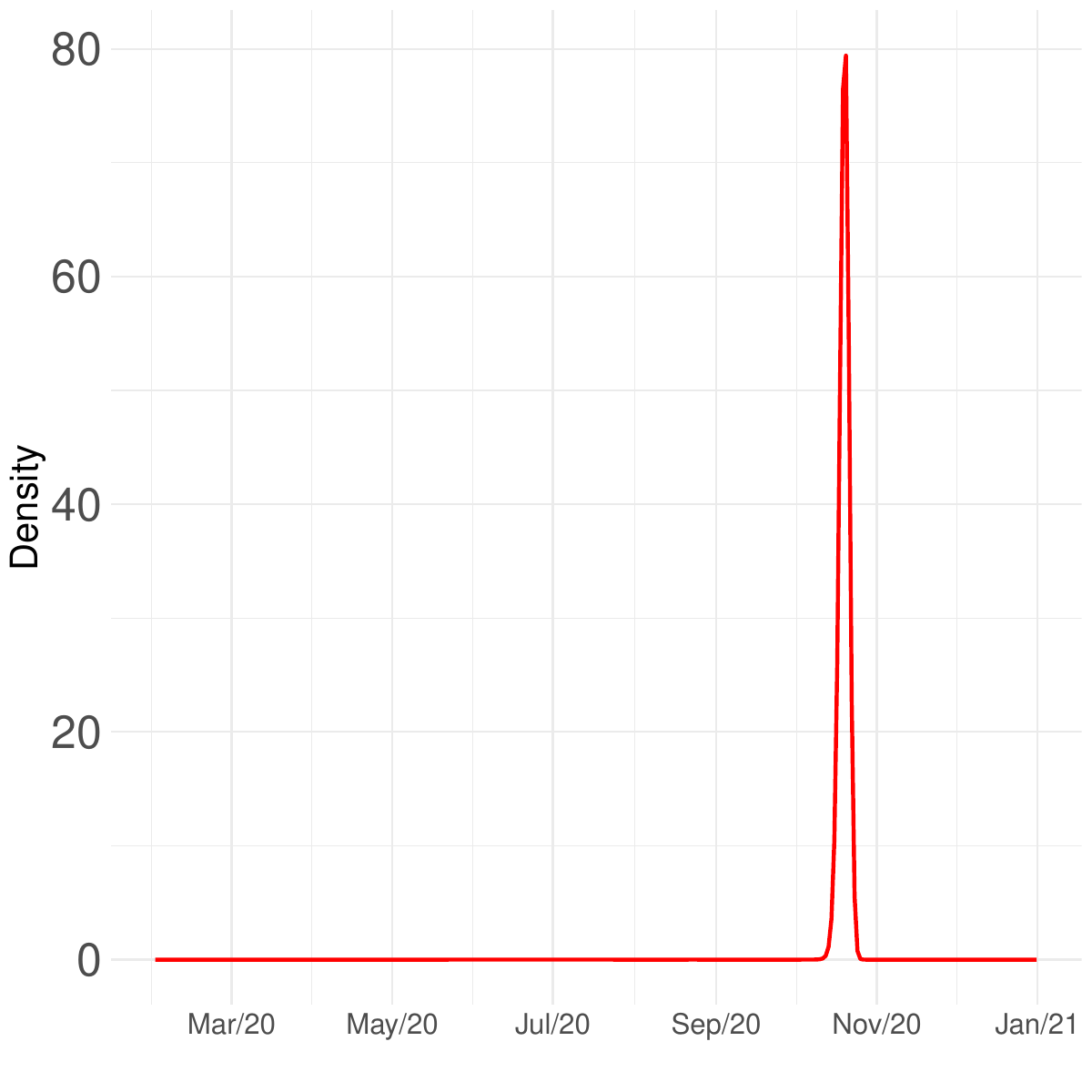}
    }
        \subfigure[Mortality rate in NL]{
        \includegraphics[width=0.4\textwidth]{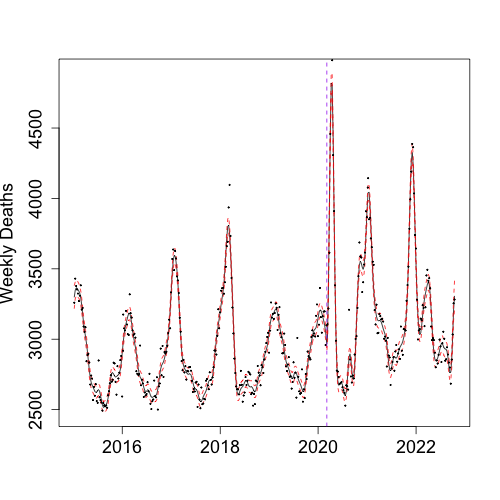}
    }
        \subfigure[Mortality rate in BG]{
        \includegraphics[width=0.4\textwidth]{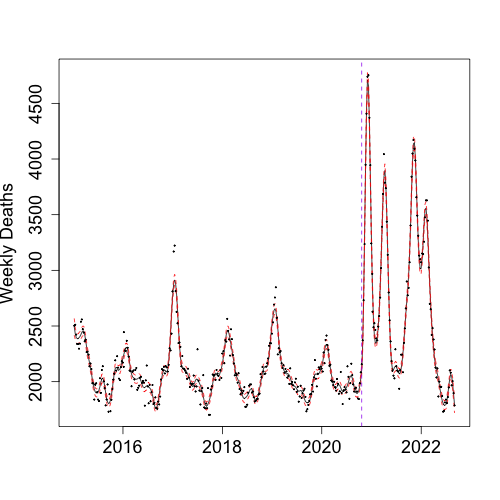}
    }
    \caption{Results for the mortality example in \cref{subsec:mortality}: (a-b) show the normalized BOSS surrogate $\Tilde{\pi}_{\texttt{BO}}(\alpha|\boldsymbol{y})$ for the Netherlands (NL) and Bulgaria (BG); (c-d) show the posterior of $\mu$ in NL and BG, with posterior mean shown in black solid line, 95 percent credible interval in red dashed lines, and observations in black dots. The change point is fixed at the posterior mode (purple dashed).
    }
    \label{fig-covid}
\end{figure}

Here $Y(t_i)$ denotes the observed weekly mortality at time $t_i$, where the unit of $t_i$ is converted to year;
the conditioning parameter $\alpha$ denotes the change-point of the mortality dynamic, and is assigned an uniform prior between the first week of the year 2019 and the first week of the year 2022.
We decompose the dynamic of mortality rate into the smooth long-term trend ($g_{tr,\text{pre}}$ or $g_{tr,\text{pos}}$) and the seasonal component ($g_{s,\text{pre}}$ or $g_{s,\text{pos}}$) with yearly variation. 
The trends are modeled using independent Integrated Wiener processes ($\text{IWP}_2$) with order 2 \citep{IwpOsplines} and the seasonal components are modeled using independent seasonal Gaussian processes ($\text{sGP}) $ with yearly periodicity and four harmonic terms \citep{sgp}.
The boundary conditions of IWP or sGP are fixed such that neither the mortality rate $\mu$ nor its derivative will have a discontinuity at the change point $\alpha$.
For priors on the variance parameters, we put independent exponential priors on their corresponding five years predictive standard deviations, with prior median $0.01$ for $g_{tr,\text{pre}}$ and $g_{s,\text{pre}}$, and median $1$ for $g_{tr,\text{pos}}$ and $g_{s,\text{pos}}$.
To facilitate the computations, all of the IWP and sGP are approximated by their finite element (FEM) approximations.

The proposed BOSS algorithm is then applied with $B = 30$ BO iterations, with the change point $\alpha$ being the conditioning parameter. 
The posterior of $\alpha$ is provided in \cref{fig-covid}, which shows that the change point is most likely at March 6th 2020 in the Netherlands, and October 20th 2020 in Bulgaria. 
Our findings are in line with the observations made in \citep{ylli2022covid}. 
They suggested that during the first half of 2020, Eastern European countries successfully implemented public health measures, resulting in a negligible excess mortality rate. 
This contrasts with most Western European countries, where the excess mortality rate exceeded expected levels by 15 to 35 percent. 
However, during the autumn/winter wave of 2020, the excess mortality rate in Eastern Europe surpassed that of Western Europe.
Since the posterior of the change point $\alpha$ is highly concentrated at the mode, we proceed with an empirical Bayes approach to obtain the posterior of the mortality rate $\mu(t)$ in each country, with the change point $\alpha$ fixed at their posterior modes. 
The posterior summaries of $\mu(t)$ are displayed in \cref{fig-covid}.

\subsection{Decomposition of CO2 Variation}\label{subsec:co2}

In this section, we apply the BOSS algorithm to analyze the atmospheric carbon dioxide (CO2) concentration data, which were collected from an observatory in Hawaii monthly before May 1974, and weekly afterward. 
In total, this dataset consists of $n = 2267$ observations.

Existing work have established clear evidence that there exists a 3-5 years cyclic variation in the CO2 concentration, related to the climate cycle. 
For example, \cite{bacastow1980atmospheric} reported correlation between the changes in atmospheric CO2 concentration and El Nino and the southern oscillation index (ENSO), which had an approximately 4-year periodicity. 
With the same dataset, \cite{rust1979inferences} later also identified a 44-month cycle which seemed to associate with ENSO. 
However, it is challenging to precisely know the exact periodicity of ENSO in the CO2 variation. 
In this section, we aim to illustrate the practical usage of the proposed BOSS algorithm, by fitting a Bayesian hierarchical model with a cyclic component that has an unknown periodicity. 
This component is introduced to capture the cyclic variation in the CO2 concentration that is likely associated with ENSO.
As a result, our model will provide (i) an inference of the periodicity of ENSO in the CO2 variation and (ii) a decomposition of the CO2 variation with full uncertainty quantification.

\begin{figure}
          \centering
          \subfigure[Posterior of $\alpha$]{
      \includegraphics[width=0.41\textwidth]{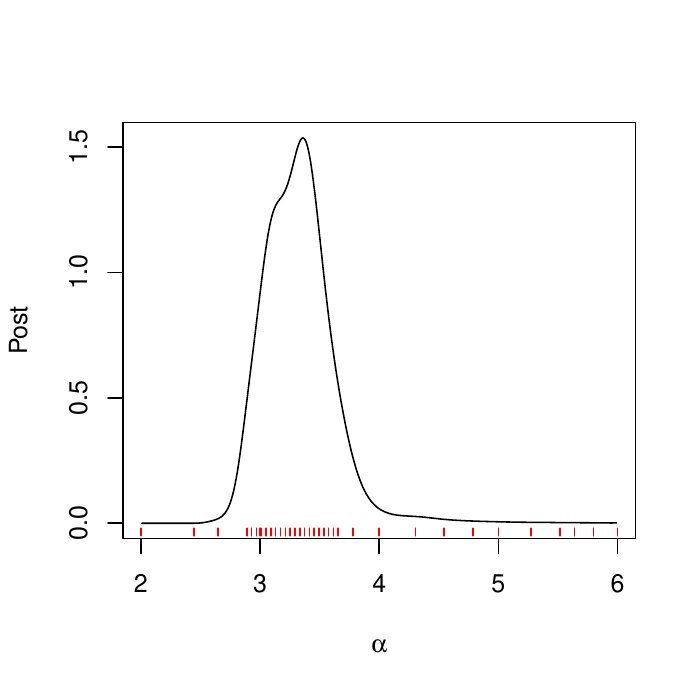}
    }
          \subfigure[Posterior of $g_{tr} + g_{1} + g_{\frac{1}{2}} + g_{\alpha}$]{
      \includegraphics[width=0.41\textwidth]{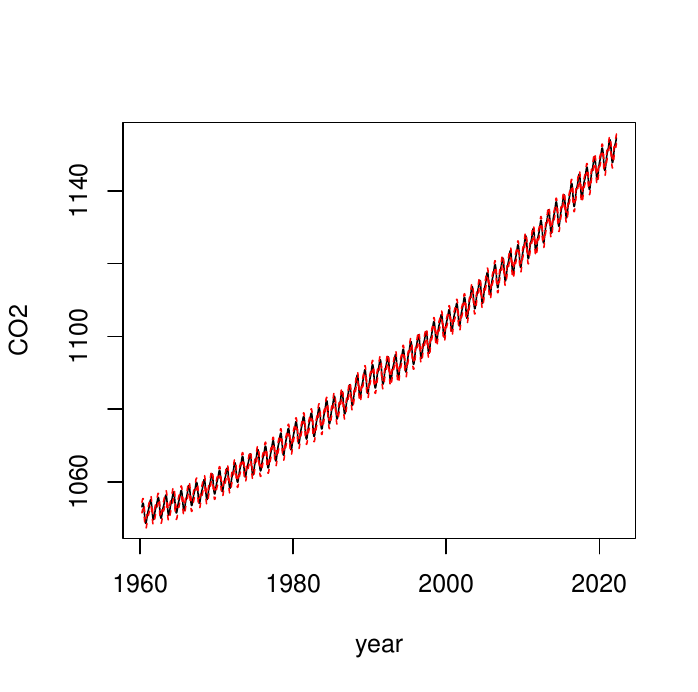}
    }
              \subfigure[Posterior of the trend $g_{tr}$]{
      \includegraphics[width=0.41\textwidth]{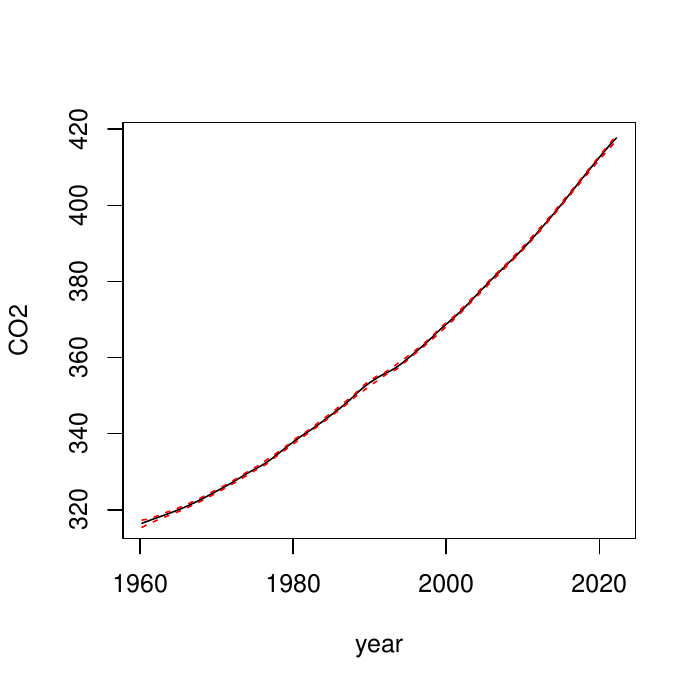}
    }
          \subfigure[Posterior of $g_{1} + g_{\frac{1}{2}} + g_{\alpha}$]{
      \includegraphics[width=0.41\textwidth]{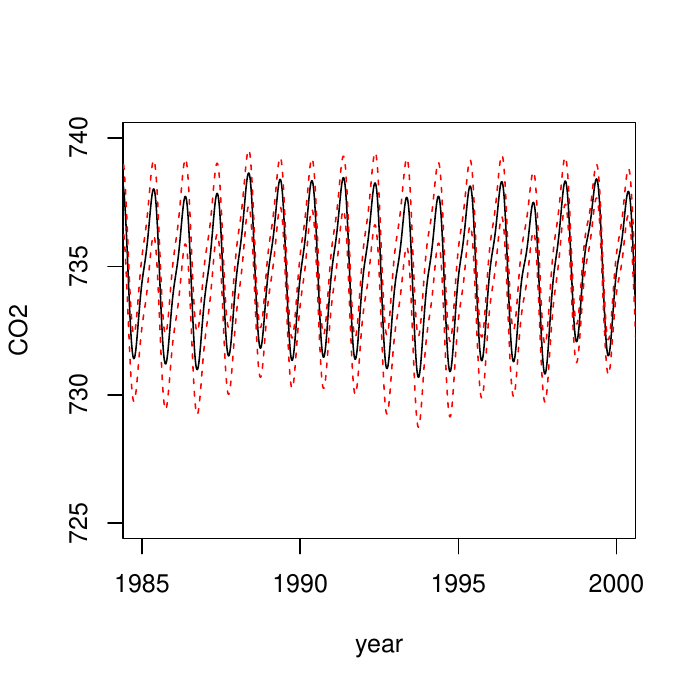}
    }
    \caption{Posteriors for the CO2 example in \cref{subsec:co2}: (a) shows the normalized BOSS surrogate $\Tilde{\pi}_{\texttt{BO}}(\alpha|\boldsymbol{y})$ (black line) and the selected BO design points (red bars); (b-d) shows the posterior median (black solid) and the 95 percent interval (red dashed). }
    \label{fig:co2}
\end{figure}

Motivated by the existing literature, the hierarchical model has the following structure:
\begin{equation}\label{equ:co2model}
    \begin{aligned}
        y_i &= g_{tr}(x_i) + g_{1}(x_i) + g_{\frac{1}{2}}(x_i) + g_{\alpha}(x_i) + e_i,\\
        g_{1} &\sim \text{sGP}_1(\sigma_{s}),\  g_{\frac{1}{2}} \sim \text{sGP}_{\frac{1}{2}}(\sigma_{s}) \\
        g_{\alpha} &\sim \text{sGP}_{\alpha}(\sigma_{\alpha}),\  g_{tr} \sim \text{IWP}_2(\sigma_{tr}), \\
        e_i &\sim \mathcal{N}(0,\sigma_e^2).
    \end{aligned}
\end{equation}
Here $y_i$ denote the CO2 concentration observed at the time $x_i$ in years since March 30, 1960. 
The component $g_{tr}$ represents the long term growth trend, modeled using a second order Integrated Wiener process (IWP) \citep{IwpOsplines}.
The components $g_1$ and $g_{\frac{1}{2}}$ represent the annual cycle and its first harmonic, modeled using a seasonal Gaussian process (sGP) with one-year and half-year periodicity \citep{sgp}.
The component $g_{\alpha}$ is the other cyclic component, modeled with a sGP with ${\alpha}$-year periodicity, where $\alpha$ is an unknown parameter assigned with an uniform prior between $2$ to $6$. 
All the boundary conditions of the sGP and the IWP are assigned with independent priors $\mathcal{N}(0,1000)$.

At a given value of $\alpha$, we assign the following independent exponential priors to the four variance parameters:
\begin{equation}
    \begin{aligned}
        &\mathbb{P}(\sigma_{s}(10) > 1) = 0.5,\ \mathbb{P}(\sigma_{\alpha}(10) > 1) = 0.5,\\
        &\mathbb{P}(\sigma_{tr}(10) > 30) = 0.5, \
        \mathbb{P}(\sigma_{e} > 1) = 0.5,
    \end{aligned}
\end{equation}
where the notation $\sigma(h)$ denotes the $h$-year predictive standard deviation (PSD) of the GP, which is a scaled version of $\sigma$ that has consistent interpretation across different GPs, see \cite{IwpOsplines} for more details. 
To facilitate the computation of the four independent GPs, we apply the finite element method to approximate each GP, using 40 equally spaced O-Splines for IWP \citep{IwpOsplines}, and 90 equally spaced sB-Splines for each sGP \citep{sgp}.

Once the value of $\alpha$ is fixed, the above model in \cref{equ:co2model} can be directly fitted using the approximate inference methods in \cite{inla} or \cite{elgm}, where the latent field $\big[g_{tr}(\boldsymbol{x}), g_{1}(\boldsymbol{x}), g_{\frac{1}{2}}(\boldsymbol{x}), g_{\alpha}(\boldsymbol{x}) \big]$ is Gaussian with a precision matrix controlled by the variance parameters $\boldsymbol{\theta} = \big[-2\log(\sigma_{tr}), -2\log(\sigma_{s}), -2\log(\sigma_{\alpha}), -2\log(\sigma_{\epsilon}) \big]$. 
However, since we are interested in making inference of the unknown periodicity $\alpha$, the existing approximate inference methods will not directly apply. 
As at each value of $\alpha$, different finite element approximation is needed with different sB-Spline basis. 
Therefore, both the design and the precision matrix of this model will vary in an intractable way as $\alpha$ varies.

Let $\alpha$ be the conditioning parameter, the proposed BOSS algorithm is then implemented as described in \cref{sec:method}, with $5$ starting values equally spaced in $[2,6]$ and $30$ BO iterations. 
The (normalized) BOSS surrogate of $\Tilde{\pi}_{\texttt{BO}}(\alpha|\boldsymbol{y})$ is shown in \cref{fig:co2}. 
The posterior mode of $\Tilde{\pi}_{\texttt{BO}}(\alpha|\boldsymbol{y})$ is found at $3.36$-years, with the $50\%$ credible interval being $[3.11, 3.47]$, and the $95\%$ credible interval being $[2.82, 3.82]$. 
This shows mild evidence toward a slightly smaller periodicity than the 44-month (3.67-year) periodicity identified in \cite{rust1979inferences}. 
Finally, to examine the overall CO2 concentration dynamic, as well as its decomposition into the trend and seasonal components, we apply the surrogate AGHQ method described in \cref{sec:method} with $10$ quadrature points. 
The posterior results of the overall dynamic as well as the decomposition can also be found in \cref{fig:co2}. 

\subsection{Detecting Ultra-Diffuse Galaxies}

In this section, we consider using our BOSS algorithm to detect Ultra-Diffuse Galaxies \citep[UDGs;][]{VanDokkum2015}. UDGs are a class of recently discovered galaxies that are of extremely low brightness, hence difficult to be detected in astronomical images \citep{Abraham2014, VanDokkum2015}. 

Recently, \cite{Li_2022, li2024poisson} entertained the idea of detecting UDGs by searching the clustering signals of their constituent star clusters. Star clusters appear as bright point sources in astronomical images, and they will not cluster together if they are not associated with a galaxy that provides gravitational potential to bind them together. Thus, if one finds clustering signals of groups of star clusters that do not belong to any observable normal, bright galaxy, the clustering signals then likely indicate the presence of UDGs. 

\cite{Li_2022} modeled the observed locations of star clusters in an astronomical image $S \subseteq \mathbb{R}^2$ as an inhomogeneous Poisson process (IPP) $\mathbf{X} \subseteq S$, and proposed the following log-Gaussian Cox process \citep[LGCP;][]{Moller1998} to detect UDGs:
\[\label{UDG LGCP}
\mathbf{X} &\sim \mathrm{IPP}(\lambda(s)),\\
\log(\lambda(s)) &= \beta_0 + \sum_{k=1}^{N_G}\beta_k\exp\left[-\left\{\frac{r(s; c_k)}{R_k}\right\}^{1/n_k}\right] + \mathcal{U}(s),\\
\mathcal{U}(s) &\sim \mathcal{GP}(\mathbf{0}, \sigma^2\text{Mat\'{e}rn}(\cdot; h, \nu)), \\
\text{Cov}(\mathcal{U}(s), \mathcal{U}(t)) &= \text{Mat\'{e}rn}(|s-t|; h, \nu), \ s, t \in S.
\]
In the above model, \citeauthor{Li_2022} used $\beta_0$ to model the intensity of star clusters that belong to the ``background" or the intergalactic space. The terms $\exp\left[-\left\{\frac{r(s; c_k)}{R_k}\right\}^{1/n_k}\right], \ k = 1, \dots, N_G$ are non-linear covariate effects used to model the intensity of star clusters in $N_G$ number of normal, bright galaxies observed in $S$. $r(s; c_k)$ is the distance from $s \in S$ to the known center $c_k$ of the $k$-th bright galaxy. $R_k$ is the characteristic size of the star cluster system of the $k$-th bright galaxy, while $n_k$ determines how the intensity of star clusters decreases as one moves away from $c_k$. $\mathcal{U}(s)$ is a zero-mean spatial Gaussian process with a Mat\'{e}rn covariance function ($\nu = 1$) that captures any unexplained clustering signals in the star cluster point patterns. The unexplained clustering signals are attributed to the existence of UDGs and they are detected by computing the exceedance probability based on the posterior distribution of $\mathcal{U}(s)$.

Due to the presence of the nuisance parameters $\{(R_k, n_k)\}_{k=1}^{N_G}$, \cref{UDG LGCP} is not a LGM or ELGM. \citeauthor{Li_2022} used the \texttt{inlabru} package which approximated the non-linear covariate effect by a first-order Taylor expansion. Inference is then conducted using INLA based on the approximated model. We here fit the model in \ref{UDG LGCP} using our proposed BOSS algorithm with the conditioning $\boldsymbol{\alpha} = (R_1, ..., R_{N_G}, n_1, ..., n_{N_G})$ to one of the star cluster data set analysed by \citeauthor{Li_2022} and obtained by \cite{Harris2020} through the \textit{Hubble Space Telescope}. We then compare our results to ones obtained by \texttt{inlabru}. The data consist of the imaged locations of $109$ star clusters. The image $S$ is roughly a square with $76$~kpc per side. Furthermore, the image is validated against a known catalogue of UDGs where it contains one confirmed UDG and one normal, bright galaxy. Thus, $N_G = 1$ and we write $\boldsymbol{\alpha} = (R, n)$.

For the prior distributions, we set
\[
\beta_0 \sim \mathcal{N}(0, 1000), \ & \beta_1 \sim \mathcal{N}(0, 1000), \\
R \ (\text{kpc}) \sim \text{Log-Normal}(2.324, 0.25^2), \  & n \sim \text{Log-Normal}(0, 0.3^2), 
\]
and the following penalized complexity (PC) priors \citep{pcprior} are chosen for the hyper-parameters $h$ and $\sigma$ of the latent Gaussian random field:
\[
 \mathbb{P}(h < 7 \text{~kpc}) = 0.5, \ \mathbb{P}(\sigma > 1.5) = 0.5.
\]

\begin{figure}
    \centering
    \subfigure[Joint posterior distribution of $(R, n)$ and posterior mode from BOSS algorithm]{\includegraphics[width = 0.46\textwidth]{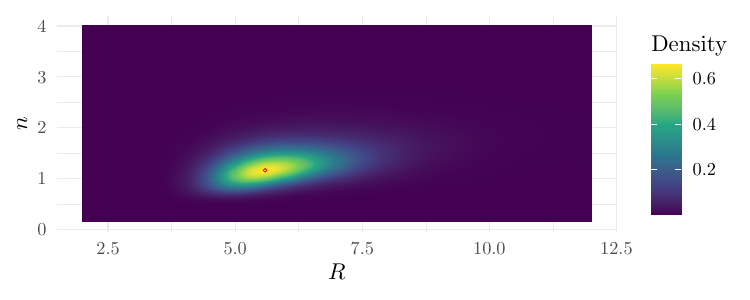}}
    \subfigure[Joint posterior distribution of $(R, n)$ and posterior mode from \texttt{inlabru}]{\includegraphics[width = 0.46\textwidth]{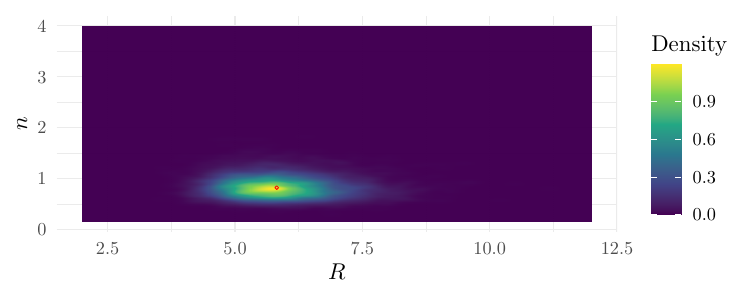}}
    \caption{Joint posterior distribution of $(R, n)$ obtained by (a) our BOSS algorithm and (b) \texttt{inlabru}. Red circles are the respective posterior mode obtained from our BOSS algorithm and \texttt{inlabru}.}
    \label{fig:joint_post_R_n}
\end{figure}

\begin{figure}
    \centering
    \subfigure[Marginal posterior of $R$]{\includegraphics[width = 0.45\textwidth]{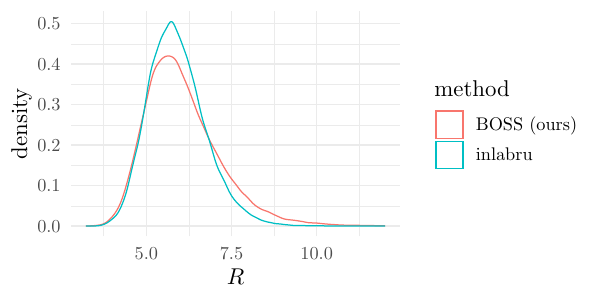}}
    \subfigure[Marginal posterior of $n$]{\includegraphics[width = 0.45\textwidth]{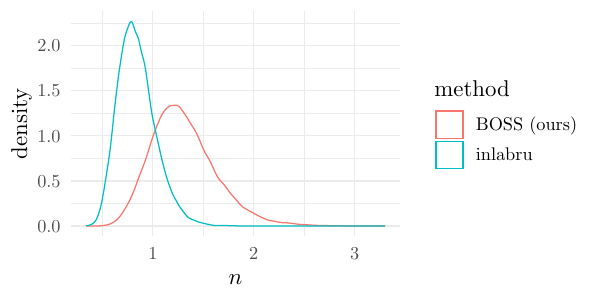}}
    \caption{Marginal posterior distribution of $R$ and $n$ from our BOSS algorithm (red) and \texttt{inlabru} (blue).}
    \label{fig:UDG_marginal}
\end{figure}

We use $100$ BO iteration in the BOSS algorithm with the parameter search space being $R \in [2, 12]$ and $n \in [0.15, 4]$. Figure \ref{fig:joint_post_R_n} showcases the joint posterior distributions of $(R, n)$ obtained from our BOSS algorithm and \texttt{inlabru}, while Figure \ref{fig:UDG_marginal} demonstrates the posterior marginal distribution of $R$ and $n$. 

From the two figures, we can see posterior distributions of $(R,n)$ differ significantly between the ones from our BOSS algorithm and the ones from \texttt{inlabru}. The biggest difference is between posterior marginals for $n$ of the two methods, while the difference between posterior marginals for $R$ is relatively small. Specifically, the posterior mode estimate of $n$ from \texttt{inlabru} is much lower than our estimate. 

To investigate, we compute the value of the objective function, i.e., the unnormalized posterior $\pi(R,n\mid \mathbf{X})$, at the respective posterior mode obtained by our BOSS algorithm and the one from \texttt{inlabru}. The objective function evaluates to $-847.0105$ at the posterior mode of $(5.587, 1.160)$ from the BOSS algorithm, while the objective function returned $-848.4749$ at the posterior mode of $(5.818, 0.823)$ from \texttt{inlabru}. 
Since the numerical approximation error from the computed log-marginal likelihood using INLA is negligible compared to the difference between the above objective function values, the numerical test here shows that the posterior results for $(R,n)$ obtained by \texttt{inlabru} are inaccurate. 
The culprit is the approximation of the non-linear covariate effect through the first-order Taylor expansion: the functional form of the covariate effect in \cref{UDG LGCP} is highly non-linear, especially with respect to the inverse power parameter $n$, causing \texttt{inlabru} to provide inaccurate results.

\begin{figure}
    \centering
    \includegraphics[width = 0.8\textwidth]{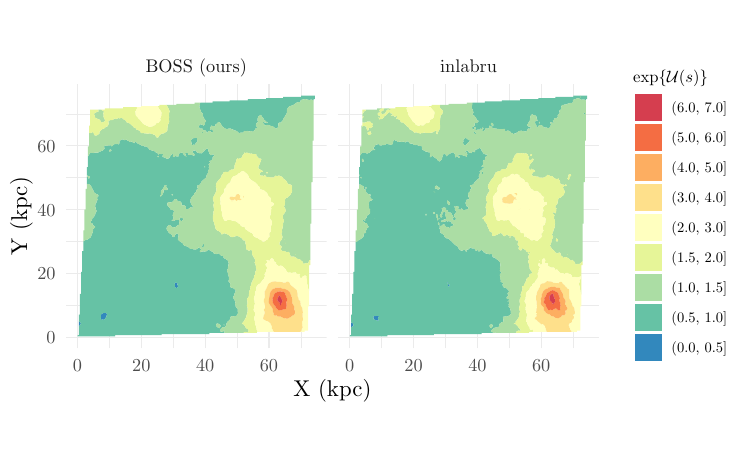}
    \caption{Posterior median of $\exp\{\mathcal{U}(s)\}$ obtained using BOSS algorithm and \texttt{inlabru}.}
    \label{fig:post_med_Us}
\end{figure}

\begin{figure}
    \centering
    \includegraphics[width = 0.8\textwidth]{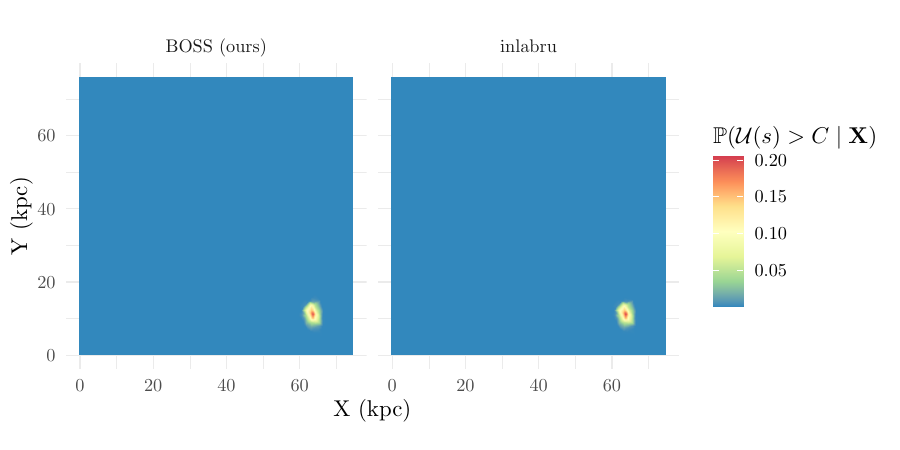}
    \caption{Posterior exceedance probability obtained by our BOSS algorithm and \texttt{inlabru}.}
    \label{fig:exceed_map}
\end{figure}

Since the goal of the original application is to detect UDGs through inspection of the posterior distribution of the spatial random field $\mathcal{U}(s)$, we also compare the posterior median of $\exp\{\mathcal{U}(s)\}$ obtained using BOSS algorithm and \texttt{inlabru} in Figure \ref{fig:post_med_Us}. Figure \ref{fig:exceed_map} shows the exceedance probability of $\mathcal{U}(s)$. The results from our BOSS algorithm in Figure \ref{fig:post_med_Us} and \ref{fig:exceed_map} are obtained using an $4$-th order adaptive quadrature through AGHQ (total of $4^2 = 16$ quadrature points). For the exceedance probability, we followed the procedure in \cite{Li_2022} and set the excursion threshold $C$ to the posterior marginal $1 - \alpha$ quantile $Q_{\alpha}$ of $\mathcal{U}(s)$. In Figure \ref{fig:exceed_map}, we have set $C = Q_{0.005}$. The computation of the exceedance probability follows \cite{Bolin2015, Bolin2018}.

The high intensity and high probability region in the lower-right corner in both Figure \ref{fig:post_med_Us} and \ref{fig:exceed_map} is the detected UDG in the considered image. Both figures show that our BOSS algorithm and \texttt{inlabru} produce results that are near identical, despite the massive difference in the inferred posterior distribution of $(R, n)$. The likely reason is that the size of the normal, bright galaxy in the considered image is quite small, thus its effect on the posterior spatial random field $\mathcal{U}(s)$ is insignificant. 

The results shown here demonstrate that the performance of our BOSS algorithm is comparable to \texttt{inlabru} in the specific problem of inferring the posterior spatial random field and the exceedance probability. However, if the goal of the application shifts  towards inferring the posterior distribution of the nuisance parameters $(R,n)$, the inference using the BOSS algorithm is then significantly more accurate than the inference using \texttt{inlabru}, as reflected by the above numerical investigation.

\section{Discussion}\label{sec:discussion}

In this paper, we propose the Bayesian Optimization Sequential Surrogate (BOSS) method for efficient approximate Bayesian inference in conditional latent Gaussian models (LGMs). Leveraging the Bayesian optimization (BO) algorithm, BOSS strategically selects design points that encapsulate the majority of the posterior mass, with the evaluation of each unnormalized density performed with efficiency through modern posterior approximation techniques. A surrogate for the unnormalized posterior density is sequentially constructed based on selected design points and their evaluations, facilitating its normalization with minimal computational overhead. We demonstrate the superior efficiency of BOSS over traditional grid-based and MCMC-based methods through comprehensive simulation studies and highlight its wide applicability with several real-world applications.

There are several aspects of the BOSS algorithm that warrant further investigation in future studies. For instance, the development of a theoretical convergence rate for the surrogate posterior generated by BOSS remains an integral part of our ongoing research. Additionally, we employed the square exponential covariance function for implementing Bayesian optimization, updating its two parameters using a simple adaptive method detailed in the supplementary material. Exploring the optimal selection of covariance functions for Bayesian optimization, particularly under varying assumptions about the true posterior distribution, is a research direction of practical relevance. Furthermore, designing adaptive methods for updating the hyperparameters of the covariance function, informed by online learning theory, could further enhance the algorithm's efficacy. These questions are left for future research.

\clearpage

\bibliographystyle{chicago}
\bibliography{bibliography}
\appendix
\section{Bayesian Optimization Details}

In the Bayesian Optimization (BO), we choose the following square exponential covariance parameterized by the length scale ${\ell}$ and standard deviation $\sigma$:
\begin{equation}\label{equ:SE_kernel}
    C(x_1, x_2) = \sigma^2 \exp\left(-\frac{\|x_1 - x_2\|^2}{2{\ell}^2}\right).
\end{equation}

The un-normalized log density \(f\) is modeled using a Gaussian Process (GP) with the covariance function specified in Equation \ref{equ:SE_kernel}.
As the density is not yet normalized, we will center the initial evaluations of $f$ to zero, and evaluate $f$ relative to the average initial evaluations.
Therefore we assume the GP on $f$ has a zero-mean function.
Due to the infinite differentiability of this covariance function, \(f\) is mean-square differentiable to any order \citep{rasmussen2003gaussian}. 
While such a strong smoothness assumption may seem unrealistic for most physical processes \citep{stein2012interpolation}, it is deemed reasonable in our context where \(f\) represents a smooth log posterior function.

During each iteration of BO, the evaluation \(f(\bm{\alpha})\) is modeled as \(f(\bm{\alpha}) = f_{\text{true}}(\bm{\alpha}) + \epsilon\), where \(\epsilon \sim \mathcal{N}(0,\tau^2)\) accounts for approximation errors in each \(f\) evaluation and aids in regularizing the matrix inversion process. 
Based on the theoretical guarantee provided in \cite{aghqtheory}, $\tau$ should be very small if each evaluation of the density is based on the use of adaptive Gauss Hermite quadrature (AGHQ), hence we choose a default of $\tau^2 = 1\times10^{-6}$.

In practice, the performance of Bayesian Optimization (BO) and, consequently, the BOSS method will depend on the choices of the hyperparameters \(\ell\) (length scale) and \(\sigma\) (standard deviation). 
To optimize these hyperparameters with a balance between accuracy and computational cost, we employ an adaptive updating mechanism via maximum likelihood estimation (MLE). 

Starting with initial values \(\ell^{(0)}\) and \(\sigma^{(0)}\), chosen based on heuristic criteria, and a predetermined update frequency \(B_A \in \mathcal{Z}^+\), we update \(\ell\) and \(\sigma\) to their MLEs after every \(B_A\) iterations of BO. This likelihood is based on the current set of design points \(\mathcal{X}_t = \{(\bm{\alpha}^{(i)}, f(\bm{\alpha}^{(i)})): i \in [t]\}\), where \(t\in\mathcal{Z}^+\) is a multiple of \(B_A\).

Specifically, the log-likelihood function is given by:
\begin{equation}
    \begin{aligned}
        \log \mathcal{L}(\ell, \sigma^2; \mathcal{X}_t) = -\frac{1}{2} \boldsymbol{f}^T (\mathbf{C} + \tau^2I)^{-1} \boldsymbol{f} - \frac{1}{2}\log \det(\mathbf{C} + \tau^2I) - \frac{t}{2} \log(2\pi),
    \end{aligned}
\end{equation}
where $\mathbf{C}$ denotes the covariance matrix constructed from evaluating the covariance function over the design points $\boldsymbol{\bm{\alpha}} = (\bm{\alpha}^{(1)}, \ldots, \bm{\alpha}^{(t)})$, and $\boldsymbol{f} = (f(\bm{\alpha}^{(1)}), \ldots, f(\bm{\alpha}^{(t)}))$ represents the vector of function evaluations at these points.
The hyperparameters $\ell$ and $\sigma$ affects the log-likelihood through the matrix $\mathbf{C}$, and their MLEs are chosen based on a fine grid search.
By default, we chose $B_A = 10$, meaning the MLEs will be updated every $10$ iterations of BO.

\end{document}